\documentclass[journal, 12pt, draftclsnofoot,onecolumn]{IEEEtran}
\ifCLASSINFOpdf
   \usepackage[pdftex]{graphicx}
   \usepackage{subfigure}
\else
\fi
\usepackage{amsmath,dsfont}
\usepackage{amssymb, color}
\usepackage{mathtools}
\usepackage{rotating}
\usepackage{cite}
\usepackage{url}

\allowdisplaybreaks[1]
\begin{document}

\newtheorem{Theorem}{Theorem}
\newtheorem{Lemma}{Lemma}
\newtheorem{Corollary}[Theorem]{Corollary}
\newtheorem{Definition}{Definition}
\newtheorem{prop}[Theorem]{Proposition}
\newtheorem{Remark}{Remark}[section]
\newtheorem{example}{Example}
\newcommand{\blue}[1]{\textcolor{blue}{#1}}

\title{Coded Caching with Heterogeneous Cache Sizes and Link Qualities: The Two-User Case}

% author names and affiliations
% use a multiple column layout for up to three different
% affiliations
\author{Daming Cao, Deyao Zhang, Pengyao Chen, Nan Liu, Wei~Kang, \\ and Deniz G\"{u}nd\"{u}z%
\thanks{D. Cao and W. Kang are with the Information Security Research Center,
Southeast University, Nanjing, China (email: \{dmcao,wkang\}@seu.edu.cn). Deyao Zhang, Pengyao Chen, Nan Liu are with the National Mobile Communications Research Laboratory,
Southeast University, Nanjing, China (email: \{dyzhang, chenpy, nanliu\}@seu.edu.cn). D. G\"{u}nd\"{u}z is with the Imperial College London, London SW7 2AZ, U.K. (email: d.gunduz@imperial.ac.uk).}%
\thanks{This work was partially supported by the National Natural Science Foundation of China under Grants $61571123$, $61571122$, and $61521061$, the Research Fund of National Mobile Communications Research Laboratory, Southeast University (No. 2017A03) and Qing Lan Project. The work of D. Gunduz is supported by the European Research Council (ERC) through Starting Grant BEACON (agreement 677854).}
\thanks{The material in this paper was presented in part at the IEEE International Symposium on Information Theory (ISIT) in Vail, Colorado in June 2018.}
%}
}

\maketitle

\begin{abstract}
Centralized coded caching problem is studied for the two-user scenario, considering heterogeneous cache capacities at the users and private channels from the server to the users, in addition to a shared channel.  Optimal caching and delivery strategies that minimize the worst-case delivery latency are presented for an arbitrary number of files. The converse proof follows from the sufficiency of file-index-symmetric caching and delivery codes, while the achievability is obtained through memory-sharing among a number of special memory capacity pairs. The optimal scheme is shown to exploit the private link capacities by transmitting part of the corresponding user`s request in an uncoded fashion. When there are no private links, the results presented here improve upon the two known results in the literature, namely, i) equal cache capacities and arbitrary number of files; and ii) unequal cache capacities and $N=2$ files. The results are then extended to the caching problem with heterogeneous distortion requirements. 
\end{abstract}

% no keywords

\IEEEpeerreviewmaketitle

\section{Introduction}
In their seminal paper \cite{maddah2014fundamental}, Maddah-Ali and Niesen propose a framework for coded caching and delivery to exploit the cache memories available at user devices to relieve the traffic burden at peak traffic periods. They consider a server holding $N$ files of equal size, serving $K$ users, each equipped with a local cache memory sufficient to store $M$ files. Users' caches are proactively filled before they reveal their demands, called the \textit{placement phase}, over a low-traffic period. In the ensuing \textit{delivery phase}, each user requests a single file from the library, which are delivered simultaneously over an error-free shared link. The coded caching scheme proposed in \cite{maddah2014fundamental} creates \emph{multicasting} opportunities by jointly designing the content placement and delivery, resulting in a global caching gain. 
The optimal caching and delivery scheme for the general coded caching problem, in terms of the worst case delivery latency, remains open despite ongoing research efforts. While many schemes have been proposed in \cite{chen2016fundamental,ji2014caching,shanmugam2016finite,gomez2016fundamental,tian2016caching,amiri2016coded,mohammadifundamental}, and converse results are presented in \cite{maddah2014fundamental,yu2017exact,sengupta2015improved,ghasemi2017improved,tian2018symmetry}, the bounds obtained do not match in general except in some special cases, i.e., $N=K=2$ \cite{maddah2014fundamental}, $N=2$ and arbitrary $K$ \cite{tian2018symmetry}, $N=3$ and $K=2$ \cite{tian2018symmetry}. The optimal caching and delivery strategy is characterized in \cite{yu2017exact} when the cache placement is constrained to be \emph{uncoded}.

Due to the difficulty of the problem, most of the literature follows the symmetric setting of  \cite{maddah2014fundamental}, in which all the users are equipped with the same cache size, and the link between the server and the users is an error-free shared bit-pipe. However, in practice, owing to the heterogeneous nature of devices, the equal cache assumption is often not realistic. Furthermore, the delivery channel quality may be different for different users, while limiting the model to a single shared link is equivalent to targeting the user with the worst channel quality. Heterogeneous cache sizes with a shared link has been considered in \cite{wang2015fundamental,amiri2017decentralized, ibrahim2017centralized, ibrahim2018benefits, ibrahim2018coded, yang2016coded}, heterogeneous link qualities has been considered in \cite{zhang2016wireless, naderializadeh2017optimality, ibrahim2017optimization}, while a few works have studied heterogeneity in \emph{both} the cache sizes and link qualities \cite{ghorbel2016content, bidokhti2016erasure, amiri2017cache, bidokhti2016noisy, bidokhti2017benefits}. References \cite{bidokhti2016erasure, bidokhti2016noisy, bidokhti2017benefits, amiri2017cache, amiri2018broadcast, Amiri:ISIT:18} take a more general approach, and consider a broadcast channel from the server to the users during the delivery phase. These papers propose cache allocation among users with different channel qualities, where it is shown that a general rule of thumb is to assign more cache to users with weaker links. We note, however, that, the cache capacity, in practice, cannot be distributed across user devices dynamically, but rather given as a fixed parameter. For example, a mobile phone with a weak link to the server is unlikely to have a larger cache than a laptop with a stronger link. Hence, we assume that both the cache capacities and the link qualities are given, and we aim to find the best \textit{centralized} caching and delivery strategy that minimizes the worst-case delivery latency. In centralized caching, we assume that the cache and link capacities of the users that participate in the delivery phase are known in advance during the placement phase, although their particular demands are not known. Therefore, their cache contents can be coordinated in a centralized manner.

To model the heterogeneous link qualities of $K$ users we consider orthogonal common and private links from the server to the users. The multicast rate tuple is specified by $(R_{\mathcal{D}})_{D \subseteq \{1,2,\ldots,K\}}$, where $R_{\mathcal{D}}$ is the rate of the common message that can be reliably transmitted to the subset of users in $\mathcal{D}$. In practice, this might model a scenario with orthogonal error-free finite-capacity channels for each subset of users, either because an orthogonal frequency band is allocated for every subset of users, or because the underlying physical layer coding and modulation schemes that dictate these rates are fixed, and the coded caching scheme is implemented on a higher layer of the communication network stack. 

Given the cache capacities $(M_1, M_2, \ldots, M_K)$, and the multicast rate tuple $(R_\mathcal{D})_{\mathcal{D} \subseteq \{1,2,\ldots,K\}}$ for the delivery phase, we are interested in finding the optimal centralized caching and delivery scheme that minimizes the delivery latency across all demand combinations. The optimal strategy will show us how to best utilize the heterogeneous caches at the users, and what to transmit over the shared and private links for the most efficient use of the communication resources.

In this paper, we focus on the special case of $K=2$ users, while the number of files, $N$, is arbitrary. We reemphasize that the optimal solution has been open even in this limited setting. Moreover, the solution presented for this special case will provide insights into the more general problem. In particular, we characterize the optimal cache and delivery strategy for a generic scenario defined with five parameters $(M_1, M_2, R_c, R_{p1}, R_{p2})$, where $R_c$ is the rate of the common message that can be transmitted to both users, while $R_{pk}$ is the rate of the private message to User $k$, $k=1,2$. The main contributions of this paper can be summarized as:
\begin{enumerate}
    \item We provide a converse result based on Tian's observation in \cite{tian2018symmetry} that it suffices to consider \textit{file-index symmetric} caching schemes in this problem.
    
    \item For $K=2$ users with heterogeneous caches and only a shared common link, we identify the optimal cache and delivery strategy for an arbitrary number of $N \geq 3$ files. Previously, only the case of $M_1=M_2, N \geq 2$ \cite{tian2018symmetry}, and $M_1 \neq M_2$ and $N=2$ \cite{yang2016coded} cases were solved.
    
    \item For the general case with one common and two private links, we find the optimal caching and delivery strategy for $N \geq 2$ files. We show that: i) the private links are used to transmit part of the requested files in an uncoded fashion; ii) for the user with the smaller-capacity private link, part of the request will be transmitted over the shared common link in an uncoded fashion unless that part of all the files are cached in the said user's cache.
    
    \item By identifying the parallels between the coded caching problem with one common and two private links studied here, and the coded caching problem with heterogeneous distortion requirements studied in \cite{yang2016coded} for the case of $K=2$ users with heterogeneous caches, we prove the optimal caching and delivery strategy also for that problem for $N \geq 3$ files. In \cite{yang2016coded}, the optimal cache and delivery strategy is characterized only for $N=2$. 

\end{enumerate}

\subsection{Notations}
Throughout this paper, for $ n \in \mathds{Z}^+$, $[n]$ denotes the index set $\{1,2,\dots,n\}$. Entropy $H(X)$ and mutual information $I(X;Y)$ are defined in the standard way.

\section{System Model} \label{Sys_Mod}
We consider a coded caching problem with one server connected to $K=2$ users.  The server has access to a database of $N$ independent equal-size files, each consisting of $F$ bits, denoted by $W_{1},W_{2},\ldots,W_{N}$. Both users are equipped with local caches, with capacities of $M_1F$ and $M_2F$ bits, respectively. The system operates in two phases. In the \textit{placement phase}, the users are given access to the entire database and fill their caches in an error-free manner. The contents of the caches after the placement phase are denoted by $Z_1$ and $Z_2$, respectively. In the delivery phase, each user requests a single file from the server, where $d_k$ denotes the index of the file requested by User $k$, $k=1,2$. After receiving the demand pair $D \triangleq (d_1,d_2)$, the server transmits messages over the available shared and private channels to the two users to satisfy their demands. 

In \cite{maddah2014fundamental} and most of the following literature, the delivery channel is modeled as an error-free shared link of limited capacity. However, in practice, the channels between the server and the users are typically of different quality. Thus, we model the delivery channel as consisting of two private error-free links with capacities $R_{p1}F$ and $R_{p2}F$ bits per unit time to User $1$ and User $2$, respectively, in addition to a shared link of capacity $R_cF$ bits per unit time.

%We model each file as a uniformly random random variable from the set $[2^{nF}]$\footnote{For $X \in \mathbb{R}^+$, $[X]$ denotes the set $\{1, \ldots, \ceil{X}\}$.}, where $F$ denotes the rate of each file per use of the delivery channel, and $n$ denotes the number of uses of the delivery channel. The server can reliably transmit an index from the set $2^{nR_pk}$ to User $k$ over the private channel, and an index from the set $2^{nR_c}$ over the common channel. 

A \textit{caching and delivery code} for this system consists of
\begin{enumerate}
    \item two caching functions
\begin{align}
\phi_k: [2^{F}]^N \rightarrow [2^{M_kF}], \quad k=1,2, \nonumber
\end{align}
which map the database into cache contents of the users, denoted by $Z_k=\phi_k(W_1, W_2, \cdots, W_N)$, $k=1,2$.

    \item $N^2$ encoding functions, one for each demand pair,
\begin{align}
f^D:  [2^{F}]^N \rightarrow
[2^{r_c^D F}] \times  [2^{ r_{p1}^D F}] \times [2^{ r_{p2}^D F}], \nonumber
\end{align}
that map the files to the messages transmitted over the common and private links, denoted as $X_c^D$, $X_{p1}^D$ and $X_{p2}^D$, respectively, i.e., $(X_c^D, X_{p1}^D, X_{p2}^D) \triangleq f^D(W_1, W_2, \cdots, W_N)$.
   
    \item $2 N^2 $ decoding functions, one for each demand pair,
\begin{align}
g_k^D: [2^{M_kF}] \times [2^{ r_c^D F}] \times [2^{ r_{pk}^D F}] \rightarrow [2^{ F}], k=1,2,  \nonumber
\end{align}
which decodes the desired file $W_{d_k}$ as $\hat{W}_{d_k}$ at User $k$ from the cached content at User $k$, the messages transmitted over the shared link and the private link to User $k$, $k=1,2$.
\end{enumerate}
The performance of a given caching and delivery code is measured by the worst-case delivery latency, which is defined as $T=\max_D T^D$, where $T^D \triangleq \max\{T^D_c,T^D_{p1},T^D_{p2}\}$,
and $T^D_c \triangleq \frac{r_c^D}{R_c}$, $T^D_{pk} \triangleq \frac{r^D_{pk}}{R_{pk}}$, $k=1,2$. In other words, $T^D$ is the latency, under demand $D$, it takes for $X_c^D$ to be received by both users while $X_{pk}^D$ is received by User $k$,  $k=1,2$.

%We provide a discussion on the worst case as follows. 
Following the idea of symmetry in \cite[Section 3]{tian2018symmetry}\cite[Definitions 3 and 4]{tian2014Characterizing}, we will exploit the symmetry among the file indexes to simplify the proof of converse. Let $\pi(\cdot)$ be a permutation function on the file index set $\{1,2,\cdots,N\}$, $\mathcal{Z}$ a subset of $\{Z_1,Z_2\}$, $\mathcal{W}$ a subset of $\{W_1,W_2,\cdots,W_N\}$, and $\mathcal{X}$ a subset of $\{X_c^D,X^D_{p1},X^D_{p2},D\in [N]\times[N]\}$. The mapping $\pi(\mathcal{W})$ is denoted by $\{W_{\pi(i)},W_i\in \mathcal{W}\}$ and the mapping $\pi(\mathcal{X})$ is denoted by $\{X_{(\cdot)}^{(\pi(d_1),\pi(d_2))},X_{(\cdot)}^{(d_1,d_2)}\in\mathcal{X}\}$. We define the \textit{file-index-symmetric codes} as follows.
\begin{Definition}
A caching and delivery code is called \textit{file-index-symmetric} if for any permutation function $\pi(\cdot)$, any subset of caches $\mathcal{Z}$, any subset of files $\mathcal{W}$, and any subset of transmitted messages $\mathcal{X}$, the following relation holds:
\begin{equation}
H(\mathcal{W},\mathcal{Z},\mathcal{X})=H(\pi(\mathcal{W}),\mathcal{Z},\pi(\mathcal{X})).
\end{equation}
\end{Definition}
Similarly to the argument on the existence of symmetric codes in \cite[Proposition 1]{tian2018symmetry}, we have the following lemma for the above problem.

\begin{Lemma}
\label{tiansym}
For any caching and delivery code, there exists a file-index-symmetric caching and delivery code with an equal or smaller worst-case delivery latency.
\end{Lemma}\par

\begin{IEEEproof}
The proof follows similar steps to the one in \cite[Proposition 1]{tian2018symmetry}. Intuitively, if we reorder the files and apply the same encoding function, the transmissions can also be changed accordingly to accommodate the requests, and it will lead to a new code that is equivalent to the original one. The proof can be completed by using a simple memory-sharing argument for these new codes.
\end{IEEEproof}

%\textit{File-index-symmetric codes} are those for which the permutations of the file index set does not lead to variations in the entropies \cite{tian2018symmetry}.

%Lemma \ref{tiansym} states that, it suffices to consider only file-index-symmetric caching and delivery codes. 
File-index-symmetric caching and delivery codes have the following property: for any pair of distinct demands $(d_1, d_2)$, i.e., $d_1 \neq d_2$, $(r_c^D, r_{p1}^D, r_{p2}^D)$ takes the same value, denoted by $(r_c, r_{p1}, r_{p2})$; similarly, for all the cases in which the two users demand the same file, i.e., $d_1=d_2$, $(r_c^D, r_{p1}^D, r_{p2}^D)$ takes the same value, denoted by $(r^0_c, r^0_{p1}, r^0_{p2})$. We are interested in the worst-case performance; hence, for the rest of the paper, we will assume $d_1 \neq d_2$. Hence, we have
\begin{align}
T=\max \left\{ \frac{r_c}{R_c},  \frac{r_{p1}}{R_{p1}},  \frac{r_{p2}}{R_{p2}} \right\}.\label{timedef}
\end{align}
We will refer to the problem described above by $\mathcal{Q}(M_1,M_2, R_c, R_{p1}, R_{p2})$. We seek the minimum achievable worst-case delivery latency $T^*(M_1, M_2, R_c, R_{p1}, R_{p2})$ across all caching and delivery codes.

\begin{Definition}
A tuple $(M_1,M_2, R_c, R_{p1}, R_{p2},T)$ is said to be \textit{achievable} if for large enough $F$, there exists a file-index-symmetric caching and delivery code with each user correctly decoding its requested file for any demand combination, i.e., $\hat{W}_{d_k}=W_{d_k}$, $k=1,2$ for all $(d_1,d_2)\in [N]\times [N]$. The  minimum achievable worst-case delivery latency is defined as 
\begin{equation}
T^*(M_1, M_2, R_c, R_{p1}, R_{p2})=\inf\{T:(M_1,M_2, R_c, R_{p1}, R_{p2},T) ~\mbox{is achievable}\}.
\end{equation}
\end{Definition}

Note that for the problem of shared common link only, i.e., $\mathcal{Q}(M_1,M_2,R_c, 0, 0)$, the capacity $R_c$ is of no significance as $r_c=T*R_c$. Hence, minimizing $T$ for a given $R_c$ is equivalent to minimizing the data rate over the shared common link, i.e., $r_c$. As a result, we denote the problem $\mathcal{Q}(M_1,M_2,R_c, 0, 0)$ by $\mathcal{Q}^c(M_1, M_2)$, and the minimal achievable data rate over the shared common link is denoted by $r_c^*(M_1,M_2)$.\par

Since we are interested in the delivery latency, to simplify the notation in the rest of the paper,  we drop the normalization measure $F$ in the rest of the paper, where the value of $H(W_i)$ is normalized as ``$1$'', $\forall i$.

\section{Shared Link Problem $\mathcal{Q}^c(M_1, M_2)$ }\label{sec1}
We start by studying the case with heterogenous cache sizes and a shared common link only, i.e., the problem $\mathcal{Q}^c(M_1, M_2)$. For this problem, we would like to minimize the data rate over the shared common link, i.e., $r_c^*(M_1,M_2)$.

The case of $K=N=2$ has been solved in \cite{yang2016coded}, and the optimal rate is shown to be
\begin{align}
r_c^*(M_1,M_2)=\max\left\{1-\frac{M_1}{2},1-\frac{M_2}{2},2-(M_1+M_2),\frac{3}{2}-\frac{M_1+M_2}{2}\right\} . \label{hcstwo}
\end{align}
Note that \cite{yang2016coded} studied the case with heterogeneous cache sizes and distortion requirements. Thus, if we consider the special case of the problem studied in \cite{yang2016coded}, in which  the distortion requirements of the two users are the same, i.e., $D_1=D_2$, or equivalently, $r_1=r_2=1$, we obtain the problem $\mathcal{Q}^c(M_1, M_2)$, and \cite[Corollary 1]{yang2016coded} provides the result in (\ref{hcstwo}).

In the case of $K=2$ and $N \geq 3$, we provide the following optimal data rate over the shared link,  which was previously unknown.
\begin{Theorem}\label{hcs}
In the cache and delivery problem $\mathcal{Q}^c(M_1, M_2)$, when $N \geq 3$, we have:
\begin{equation}
 r_c^*(M_1,M_2)=\max \left\{1-\frac{M_1}{N},1-\frac{M_2}{N},2-\frac{3M_1}{N}-\frac{M_2-M_1}{N-1},2-\frac{3M_2}{N}-\frac{M_1-M_2}{N-1}\right\}.  \label{hcsgen}
\end{equation}
\end{Theorem}
\begin{Remark}
The special case of $M_1=M_2=M$ has been solved in \cite{tian2018symmetry}, where the achievability follows from \cite{maddah2014fundamental}, while the converse proof utilizes the symmetry of optimal codes.
\end{Remark}

\subsection{The converse proof of Theorem \ref{hcs}} The first two terms of (\ref{hcsgen}) follow from the cut-set bound \cite{maddah2014fundamental}. The third and fourth terms follow from the following lemma which will be useful throughout the paper.
\begin{Lemma}\label{chenbound}
In problem $\mathcal{Q}^c(M_1, M_2)$ with $N \geq3$, the common delivery rate $r_c$ of any achievable scheme must satisfy
\begin{align}
  NM_{i}+(2N-3)M_{j}+N(N-1)r_c&\geq 2N(N-1),\quad \forall (i,j)\in\{(1,2),(2,1)\}.\label{chenboundeq1}
%  NM_{2}+(2N-3)M_{1}+N(N-1)r_c&\geq 2N(N-1).\label{chenboundeq2}
\end{align}
\end{Lemma}
The details of the proof of Lemma \ref{chenbound} is given in Appendix \ref{Nan01}. In the following we comment on some of the proof ideas. The proof follows from the proof of Lemma \ref{tiansym} with the help of two major steps stated in the following two lemmas.
\begin{Lemma}\label{skey}
In problem $\mathcal{Q}^c(M_1, M_2)$, for file-index-symmetric caching and delivery codes, we have:
\begin{align}
H(X_c^{(1,2)}|Z_i,W_1) &\geq 1-\frac{1}{N-1}[H(Z_1|W_1)+H(Z_2|W_1)],\quad \forall i=1,2. \label{Nan12}
%H(X_c^{(2,1)}|Z_2,W_1) &\geq 1-\frac{1}{N-1}[H(Z_1|W_1)+H(Z_2|W_1)]. \label{Nan12}
\end{align}
\end{Lemma}

\begin{Lemma}\label{scachelow}
For file-index symmetric caching and delivery codes, we have
\begin{align}
  NH(Z_{i}|W_{1})&\geq (N-1)H(Z_{i}),\quad \forall i=1,2.\label{Nan14}
%  NH(Z_{2}|W_{1})&\geq (N-1)H(Z_{2}). \label{Nan14}
\end{align}
\end{Lemma}
Please note that Lemma \ref{scachelow} holds for any file-index symmetric caching code, irrespective of the problem, i.e., it holds for the more general problem of $\mathcal{Q}(M_1, M_2, R_c, R_{p1}, R_{p2})$.

As it can be seen, Lemma \ref{skey} allow us to lower bound complicated terms, such as \break $H(X_c^{(1,2)}|Z_1,W_1)$, with simpler ones, such as $H(Z_1|W_1)$, while Lemma \ref{scachelow} further lower bounds terms, such as $H(Z_1|W_1)$, with even simpler ones, such as $H(Z_1)$, which is equal to the size of the cache of User 1, i.e., $M_1$. Hence, the main aim of the two lemmas is to provide a lower bound that  depends only on the placement scheme, and is independent of the delivery scheme. The same idea appeared in \cite[Lemma 1]{yu2017characterizing}.
The proofs of Lemmas \ref{skey} and \ref{scachelow} are provided in Appendices \ref{Nan04} and \ref{Nan09}, respectively.

The converse of Theorem \ref{hcs} is completed with Lemma \ref{chenbound}.

\begin{figure}
  \centering
  % Requires \usepackage{graphicx}
  \includegraphics[width=2in]{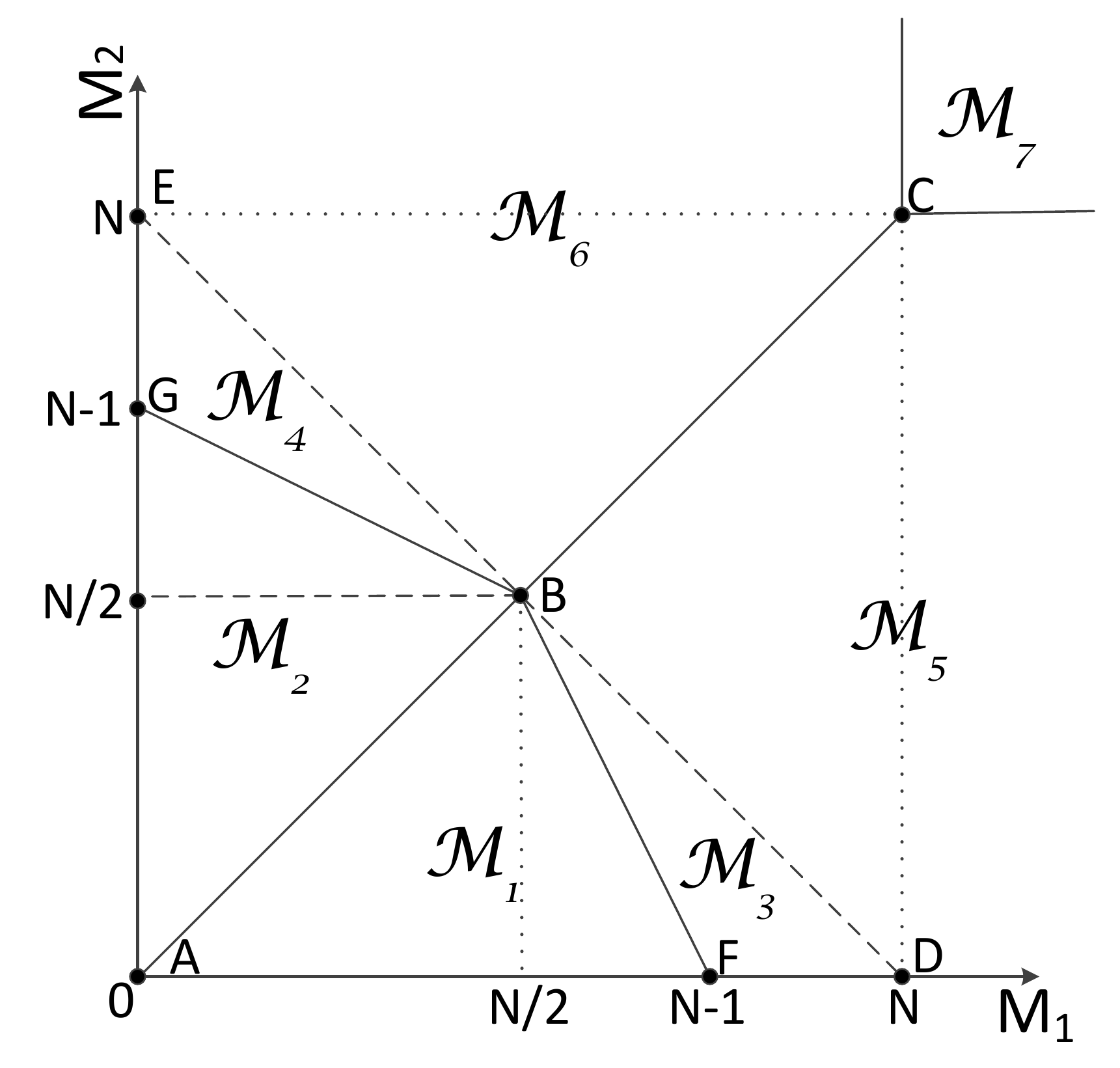}
    \caption{The optimal tradeoff between $r_c^*(M_1, M_2, R_c, 0, 0)$ and $(M_1,M_2)$ with $N\geq 3$.}\label{fenqu}
\end{figure}

\subsection{The achievability proof for Theorem \ref{hcs}} \label{Nan73}
In Figure \ref{fenqu}, we show the 2-dimensional plane of possible $(M_1,M_2)$ pairs. For the following points on this figure, the minimum data rate on the shared common link, $r_c^*$, is known:
\begin{enumerate}
\item Point A: $(M_1,M_2, r_c^*)=(0,0,2)$. This is the case with no caches at the users.
\item Point B: $(M_1,M_2,r_c^*)=(\frac{N}{2}, \frac{N}{2}, \frac{1}{2})$. This is the symmetric cache capacity scenario with the achievability proposed in \cite{maddah2014fundamental}, and its converse proved in \cite{tian2018symmetry}. The corresponding caching-delivery scheme is the following: each file is split into two parts of equal size $(W_i^1, W_i^2)$, $i=1,2,\cdots,N$. In the placement phase, User $k$ caches $\{W_i^k, i=1,2,\cdots,N\}$, $k=1,2$. The delivery scheme upon receiving request $(d_1,d_2)$ is to transmit $\{W_{d_1}^2 \oplus W_{d_2}^1\}$.
\item Point C: $(M_1,M_2, r_c^*)=(N, N, 0)$. This is the case in which the cache at each user is large enough to cache the entire library, and as such, nothing needs to be transmitted via the shared common link.
\item Point D: $(M_1, M_2, r_c^*)=(N, 0, 1)$. This is the case in which User 1 has a cache that is large enough to store the entire library, and User 2 has no cache. Thus, it is optimal to transmit only the requested file of User 2 via the shared common link.
%\item Point E: $(M_1, M_2, r_c^*)=(0, N, 1)$. This is the case symmetric to Point D by swapping the indices of 1 and 2.
\end{enumerate}

We now add the achievability scheme for Point $F$, i.e., $(M_1, M_2, r_c^*)=(N-1,0,1)$. Note that the achievability for the points symmetric with respect to the $AC$ line, i.e., points $E$ and $G$, follow directly.
\begin{itemize}
  \item \textbf{Placement phase}: User $1$ fills its cache with the module sum of every two label-adjacent files, i.e. $Z_{1}=\{W_{1}\oplus W_{2},W_{2}\oplus W_{3},\cdots,W_{N-1}\oplus W_{N}\}$.
  \item \textbf{Delivery phase}: The server transmits $X_c^{(d_{1},d_{2})}=\{W_{d_{2}}\}$. Therefore, User $2$ can directly get $W_{d_{2}}$, while user $1$ can decode $W_{d_{1}}$ with the help of its own cache by successive cancellation. For example if $(d_1,d_2)=(1,4)$, User 1 can firstly recover  $W_3$ from $(W_3 \oplus W_4, X_c^{(1,4)}=W_4)$, it then goes on to obtain $W_2$ from $(W_3, W_2 \oplus W_3)$, and finally it decodes the requested file $W_1$ from $(W_2, W_1 \oplus W_2)$.
  \end{itemize}
By performing memory-sharing\cite{maddah2014fundamental,yang2016coded,sengupta2016layered} among the seven points, i.e., Point A to Point G, we can obtain the following achievable data rate on the shared common link:
\begin{align} \label{Nan10}
r_{c}(M_1,M_2)=
\begin{cases}
2-\frac{3M_2}{N}-\frac{M_1-M_2}{N-1} & (M_{1},M_{2})\in\mathcal{M}_{1}\\
2-\frac{3M_1}{N}-\frac{M_2-M_1}{N-1} & (M_{1},M_{2})\in\mathcal{M}_{2}\\
1-\frac{M_{2}}{N} & (M_{1},M_{2})\in\mathcal{M}_{3},\mathcal{M}_{5}\\
1-\frac{M_{1}}{N} & (M_{1},M_{2})\in\mathcal{M}_{4},\mathcal{M}_{6}
\end{cases}.
\end{align}
Thus,  the achievability part of Theorem \ref{hcs} is proved.

\subsection{Comparison and analysis}
As we mentioned before, the problem $\mathcal{Q}^c(M_1, M_2)$ with $N=2$ has been solved in \cite{yang2016coded}. But for $N \geq 3$, the best known achievability schemes \cite[Section \uppercase\expandafter{\romannumeral3}-C]{yang2016coded}, \cite{sengupta2016layered}, which will be denoted as the LHC scheme here, perform memory sharing between the five points of Fig. \ref{fenqu}, i.e., Point A to Point E, and thus obtain an achievable data rate on the shared common link as
\begin{figure}
  \centering
  \includegraphics[width=2.5in]{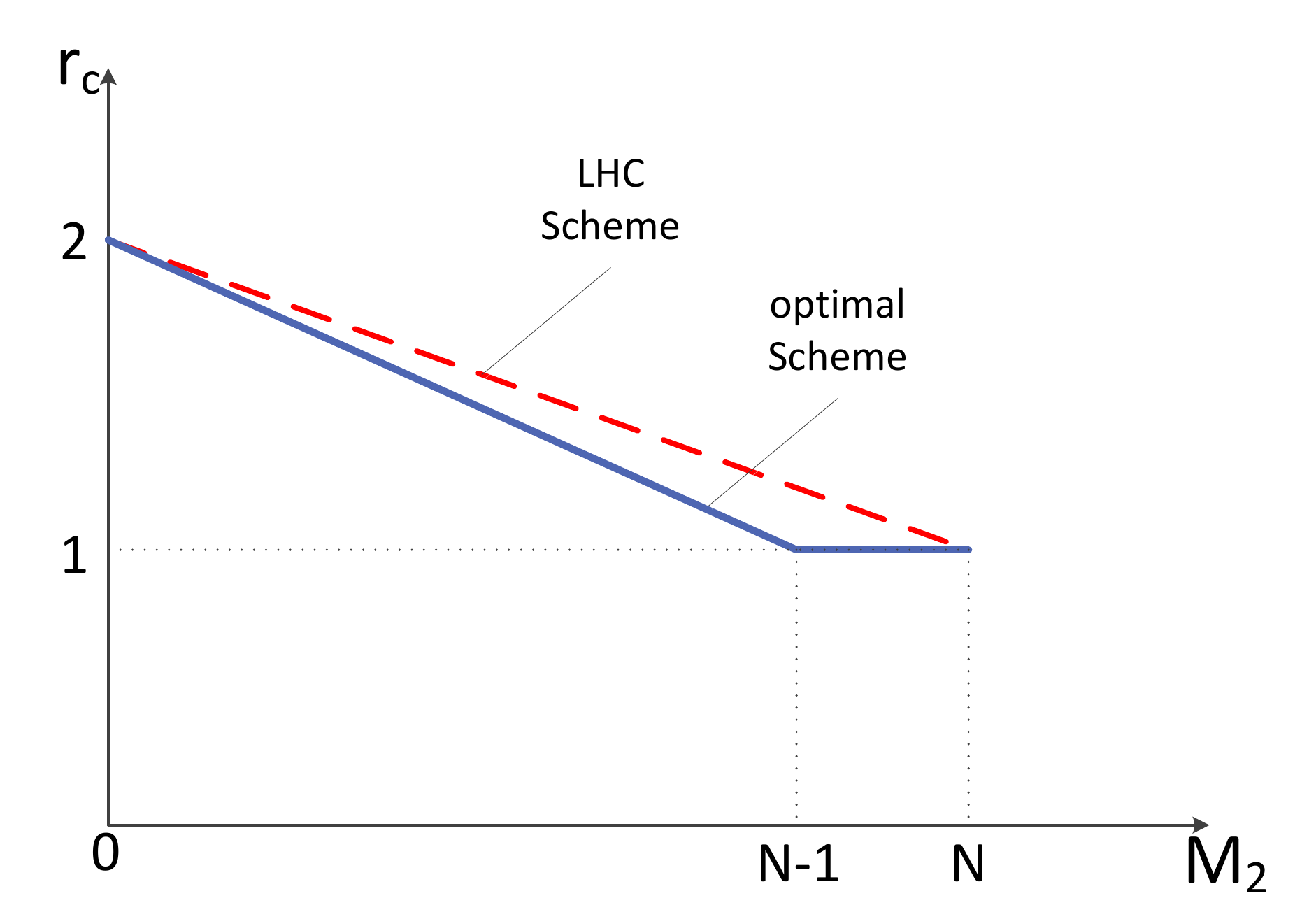}
    \caption{The comparison between our scheme and the LHC scheme for the problem $\mathcal{Q}^c(0, M_2)$}. \label{comparelayer}
\end{figure}

\begin{align}
\bar{r}_{c}(M_1,M_2)=
\begin{cases}
2-\frac{2M_{2}}{N}-\frac{M_{1}}{N} & (M_{1},M_{2})\in\mathcal{M}_{1},\mathcal{M}_{3}\\
2-\frac{2M_{1}}{N}-\frac{M_{2}}{N} & (M_{1},M_{2})\in\mathcal{M}_{2},\mathcal{M}_{4}\\
1-\frac{M_{2}}{N} & (M_{1},M_{2})\in\mathcal{M}_{5}\\
1-\frac{M_{1}}{N} & (M_{1},M_{2})\in\mathcal{M}_{6}
\end{cases}.\nonumber
\end{align}

We see that the optimal delivery rate is lower than the rate achieved by the LHC scheme, in which the delivery phase is divided into layers of unicast and multicast. We improve the delivery rate from $(M_1, M_2, r_c)=(0,N,1)$ to $(0, N-1,1)$ with the help of \emph{coded} placement. In particular, for the problem $\mathcal{Q}^c(0, M_2)$, i.e., $M_1=0$, the improvement of our scheme is plotted in Fig. \ref{comparelayer}.

\begin{figure}
  \centering
  \includegraphics[width=6in]{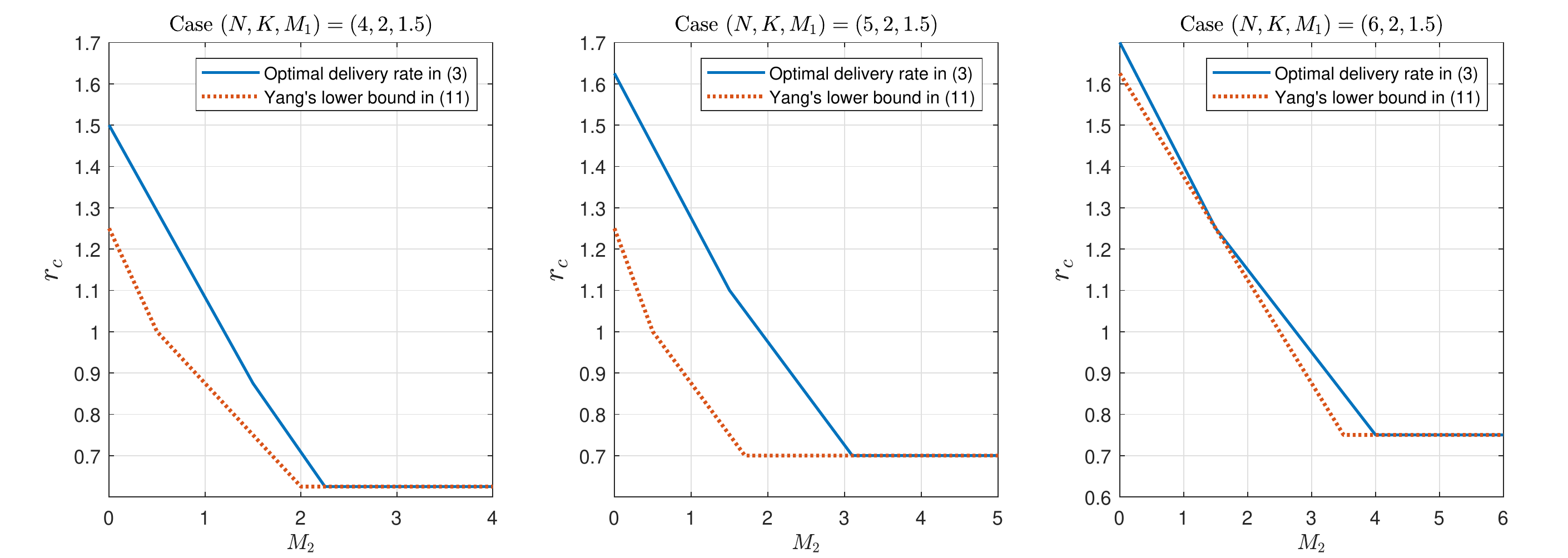}
    \caption{The comparison between our lower bound and the one in \cite{yang2016coded} (see \eqref{Deniz}) for the problem $\mathcal{Q}^c(1.5, M_2)$, for $N=4,5,6$.}\label{Nan25}
\end{figure}

As for the converse, when $N \geq 3$, the best known converse to date is given by \cite[Lemma 1]{yang2016coded}, which is the minimum of the five terms
  \begin{align}
  r_c (M_1,M_2)  \geq & \max\Bigg\{1-\frac{M_{1}}{N},1-\frac{M_{2}}{N},2-\frac{M_{1}+M_{2}}{\lfloor N/2 \rfloor}, \frac{3}{2}-\frac{M_{1}+M_{2}}{2\lfloor N/2 \rfloor},2-\frac{M_{1}+M_{2}}{2\lfloor N/3 \rfloor}\Bigg\}, \label{Deniz}
\end{align}
where the first two terms follow from the cut-set bound, the third and fourth terms follow from the straightforward generalization of the proof of the same problem for the case $N=2$. In this proof, the step \cite[Eqn.~(40c)]{yang2016coded} may be loose because the content of two caches may not be independent even conditioned on the knowledge of some files. We transform terms like $H(X_{i,j},Z_{k}|W_{i})$ into $H(X_{i,j}|Z_{k},W_{i})$ and $H(Z_{k}|W_{i})$, and then bound these two terms via Lemmas \ref{skey} and \ref{scachelow} to obtain a tighter converse. It has been argued in \cite{yang2016coded} that (\ref{Deniz}) is tight when $N$ is an integer multiple of $3$ and $M_1=M_2$. Indeed, comparing (\ref{Deniz}) and (\ref{hcsgen}), we see that when $N=3$, the two bounds are the same, which means that the bound in (\ref{Deniz}) is tight for $N=3$ and \emph{arbitrary} $(M_1, M_2)$. When $N=4, 5$ and $6$, we plot the two bounds in Fig. \ref{Nan25} to illustrate that (\ref{hcsgen}) improves upon the best known converse bound (\ref{Deniz}). Moreover, Theorem \ref{hcs} proves that (\ref{hcsgen}) is the minimum achievable data rate over the shared common link.

\section{General problem $\mathcal{Q}(M_1, M_2, R_c, R_{p1}, R_{p2})$ }
In this section, we study the general problem $\mathcal{Q}(M_1, M_2, R_c, R_{p1}, R_{p2})$, i.e., the problem with one shared common link and two private links, one for each user. We characterize the optimal delivery latency $T^*(M_1, M_2, R_c, R_{p1},R_{p2})$ in the following theorem.
\begin{Theorem}\label{hcsg}
For problem $\mathcal{Q}(M_1, M_2, R_c, R_{p1}, R_{p2})$ with $N=2$, we have:
\begin{equation}\label{hcsgtwo}
  T^*=\max\left\{\frac{1-\frac{M_1}{2}}{R_c+R_{p1}},\frac{1-\frac{M_2}{2}}{R_c+R_{p2}},\frac{2-M_1-M_2}{R_c+R_{p1}+R_{p2}} , \frac{3-M_1-M_2}{2(R_c+R_{p2})+R_{p1}},\frac{3-M_1-M_2}{2(R_c+R_{p1})+R_{p2}}\right\},
\end{equation}
while if $N \geq 3$, we have:
\begin{align}\label{hcsggen}
 T^*=&\max\Bigg\{\frac{1-\frac{M_1}{N}}{R_c+R_{p1}},\frac{1-\frac{M_2}{N}}{R_c+R_{p2}},\frac{2-\frac{3M_2}{N}-\frac{M_1-M_2}{N-1}}{R_c+R_{p1}+R_{p2}},\frac{2-\frac{3M_1}{N}-\frac{M_2-M_1}{N-1}}{R_c+R_{p1}+R_{p2}}, \nonumber\\ &\frac{N(2N-1)-2(N-1)M_{1}-NM_{2}}{N^{2}(R_c+R_{p2})+N(N-1)R_{p1}},\frac{N(2N-1)-2(N-1)M_{2}-NM_{1}}{N^{2}(R_c+R_{p1})+N(N-1)R_{p2}}\Bigg\}.
\end{align}
\end{Theorem}
\subsection{Converse proof of Theorem \ref{hcsg}} \label{Nan103}
We define $\mathcal{S}$ as the set of all possible caching and delivering codes. Then, we have
\begin{align}
T=&\min_{\mathcal{S}}\max \left\{\frac{r_{c}}{R_{c}},\frac{r_{p1}}{R_{p1}}, \frac{r_{p2}}{R_{p2}}\right\}\nonumber\\
  \geq&\min_{\mathcal{S}}\max\left\{\frac{r_c+r_{p1}}{R_c+R_{p1}},\frac{r_{p2}}{R_{p2}}\right\}\label{hcsgconeq1}\\
 % \geq&\max\min_{\mathcal{S}}\left\{\frac{r_c+r_{p1}}{R_c+R_{p1}},\frac{r_{p2}}{R_{p2}}\right\}\nonumber\\
  \geq&\min_{\mathcal{S}}\frac{r_c+r_{p1}}{R_c+R_{p1}}\nonumber\\
 % =&\frac{\min_{\mathcal{S}}\{r_c+r_{p1}\}}{R_c+R_{p1}}\nonumber\\
  \geq& \frac{1-M_1/N}{R_c+R_{p1}},\label{hcsgconeq2}
\end{align}
where (\ref{hcsgconeq1}) follows from the fact that for positive numbers $a,b,c,d,\alpha$, we have $\max \left\{ \frac{a}{b}, \frac{c}{d} \right\} \geq \frac{a+\alpha c}{b+\alpha d}$, and (\ref{hcsgconeq2}) is from the cut-set bound for User $1$. Similarly, we also have
\begin{align}
T \geq \frac{1-M_2/N}{R_c+R_{p2}}. \label{Nan47}
\end{align}

Note that any achievable scheme for problem $\mathcal{Q}(M_1, M_2, R_c, R_{p1}, R_{p2})$ can be transformed to be achievable for problem $\mathcal{Q}^c(M_1, M_2)$, because we may transmit all three signals $X_c^{(d_1,d_2)}$ with rate $r_c$,  $X_{p1}^{(d_1,d_2)}$ with rate $r_{p1}$, and $X_{p2}^{(d_1,d_2)}$ with rate $r_{p2}$, of $\mathcal{Q}(M_1, M_2, R_c, R_{p1}, R_{p2})$ over the shared common link of the problem $\mathcal{Q}^c(M_1, M_2)$, resulting in a common rate of $r_c+r_{p1}+r_{p2}$. Hence,  $r_c+r_{p1}+r_{p2}$ must satisfy Lemma \ref{chenbound}, i.e., when $N\geq 3$,
\begin{equation}
 NM_{i}+(2N-3)M_{j}+N(N-1)[r_c+r_{p1}+r_{p2}]\geq 2N(N-1),\quad \forall (i,j)\in\{(1,2),(2,1)\}. \label{hcsgchenbound}
%    NM_{2}+(2N-3)M_{1}+N(N-1)[r_c+r_{p1}+r_{p2}]&\geq 2N(N-1).\label{Nan45}
\end{equation}
Therefore, we have
\begin{align}
T=&\min_{\mathcal{S}}\max_{i=c,p1,p2}\{T_i\}\nonumber\\
\geq&\min_{\mathcal{S}} \frac{ r_c+r_{p1}+r_{p2}}{ R_c+R_{p1}+R_{p2}}\label{hcsgconeq3}\\
%=&\frac{\min_{\mathcal{S}}\{r_c+r_{p1}+r_{p2}\}}{R_c+R_{p1}+R_{p2}}\nonumber\\
\geq &\frac{\max\left\{2-\frac{3M_2}{N}-\frac{M_1-M_2}{N-1}, 2-\frac{3M_1}{N}-\frac{M_2-M_1}{N-1}\right\}}{R_c+R_{p1}+R_{p2}},\label{hcsgconeq4}
\end{align}
where (\ref{hcsgconeq3}) follows by applying twice the reasoning used for (\ref{hcsgconeq1}), and (\ref{hcsgconeq4}) follows from (\ref{hcsgchenbound}).

Note that any achievable scheme for problem $\mathcal{Q}(M_1, M_2, R_c, R_{p1}, R_{p2})$ can be transformed to be achievable for $\mathcal{Q}(M_1, M_2, R_c, R_{p1}, 0)$, because we can transmit both signal $X_c^{(d_1,d_2)}$ with rate $r_c$ and $X_{p2}^{(d_1,d_2)}$ with rate $r_{p2}$ for problem $\mathcal{Q}(M_1, M_2, R_c, R_{p1}, R_{p2})$ over the shared common link in problem \\$\mathcal{Q}^c(M_1, M_2, R_c, R_{p1}, 0)$, resulting in a rate of $r_c+r_{p2}$, while the private rate $r_{p1}$ to User 1 remaining the same. We can prove the following lemma for the problem of $\mathcal{Q}(M_1, M_2, R_c, R_{p1}, 0)$, i.e., the problem with one shared common link and one private link to User 1.
\begin{Lemma}\label{czbound}
In problem $\mathcal{Q}(M_1, M_2, R_c, R_{p1}, 0)$ with $N \geq 2$, the data rate on the shared common link $r_c$ and the only private link $r_{p1}$, must satisfy:
\begin{equation}
  N^{2}r_c+N(N-1)r_{p1} \geq N(2N-1)-2(N-1)M_{1}-NM_{2}.\label{hcspczbound}
\end{equation}
\end{Lemma}

The details of the proof of Lemma \ref{czbound}, which follows similarly to Lemma \ref{chenbound}, are relegated to Appendix \ref{Nan18}. In the proof, the following lemma, whose proof is provided in Appendix \ref{Nan22}, replaces the role of Lemma \ref{skey}.
\begin{Lemma}\label{skeycor}
In problem $\mathcal{Q}(M_1, M_2, R_c, R_{p1}, 0)$, for file-index-symmetric caching and delivery codes, we have
\begin{align}
  H(X_c^{(1,2)},X_{p1}^{(1,2)}|Z_{1},W_1) &\geq 1-\frac{1}{N-1}[H(Z_1|W_1)+H(Z_2|W_1)],\label{skeycoreq1}\\
  H(X_c^{(2,1)}|Z_{2},W_1) +r_c+r_{p1}+M_1&\geq 2+\frac{N-2}{N-1}H(Z_1|W_1)-\frac{1}{N-1}H(Z_2|W_1).\label{skeycoreq2}
\end{align}
\end{Lemma}
Again, Lemma \ref{skeycor} provides a way to lower bound terms, such as \break $H(X_c^{(i,j)}, X_p^{(i,j)}|Z_1,W_1)$, with simpler ones, such as $H(Z_1|W_1)$, and then, we again use Lemma \ref{scachelow} to lower bound terms, such as $H(Z_1|W_1)$, with simpler ones, such as $H(Z_1)$, to obtain Lemma \ref{czbound}. Thus, for problem $\mathcal{Q}(M_1, M_2, R_c, R_{p1}, R_{p2})$ with $N\geq 2$, we have
\begin{equation}
  N^{2}[r_c+r_{p2}]+N(N-1)r_{p1} \geq N(2N-1)-2(N-1)M_{1}-NM_{2}.\label{hcsgczbound}
\end{equation}
We can obtain
\begin{align}
T=&\min_{\mathcal{S}}\max_{i=c,p1,p2}\{T_i\}\nonumber\\
\geq & \min_{\mathcal{S}} \max \left\{ \frac{ r_c+r_{p2}}{ R_c+R_{p2}}, \frac{r_{p1}}{R_{p1}} \right\} \label{Nan46} \\
\geq&\min_{\mathcal{S}} \frac{ N^{2}(r_c+r_{p2})+N(N-1)r_{p1}}{ N^{2}(R_c+R_{p2})+N(N-1)R_{p1}}\label{hcsgconeq5}\\
%=&\frac{\min_{\mathcal{S}}\{N^{2}(r_c+r_{p2})+N(N-1)r_{p1}\}}{N^{2}(R_c+R_{p2})+N(N-1)R_{p1}}\nonumber\\
\geq &\frac{N(2N-1)-2(N-1)M_{1}-NM_{2}}{N^{2}(R_c+R_{p2})+N(N-1)R_{p1}},\label{hcsgconeq6}
\end{align}
where (\ref{Nan46}) and (\ref{hcsgconeq5}) follow similarly to (\ref{hcsgconeq1}); and (\ref{hcsgconeq6}) from (\ref{hcsgczbound}). By exploring the symmetry between Users 1 and 2, similarly to (\ref{hcsgconeq6}), we also have
\begin{align}
T \geq \frac{N(2N-1)-2(N-1)M_{2}-NM_{1}}{N^{2}(R_c+R_{p1})+N(N-1)R_{p2}}. \label{Nan48}
\end{align}
Hence, from (\ref{hcsgconeq2}), (\ref{Nan47}), (\ref{hcsgconeq4}), (\ref{hcsgconeq6}), (\ref{Nan48}), the proof of (\ref{hcsggen}) is completed. Note that the above upper bounds (\ref{hcsgconeq2}), (\ref{Nan47}), (\ref{hcsgconeq6}), (\ref{Nan48}) hold for any $N \geq 2$.

Finally, for the case $N=2$, we only need to prove the third term, i.e.,
\begin{align}
T^* \leq \frac{2-M_1-M_2}{R_c+R_{p1}+R_{p2}}, \nonumber
\end{align}
which follows from the cut-set bound 
\begin{align}
M_1+M_2+r_c+r_{p1}+r_{p2} \geq H(W_1,W_2)=2, \nonumber
\end{align}
and (\ref{hcsgconeq3}). Hence, the proof of (\ref{hcsgtwo}) is also complete.

\subsection{Achievability proof of Theorem \ref{hcsg} for $N \geq 3$}\label{sec:general_achieve}
The proof of achievability consists of three parts. In the first part, we find achievable schemes for a set of special points. More specifically, the achievable scheme we propose for each special point is a generalization of the achievable scheme proposed for the special point $(M_1, M_2)$ of problem $\mathcal{Q}^c(M_1,M_2)$, studied in Section \ref{Nan73}. In the second part, we  perform memory-sharing and time-sharing among the special points obtained in the first part to construct a set of achievable schemes for the current problem. In the third part, we show that there exists an achievable point $(M_1,M_2, r_c, r_{p1}, r_{p2})$ within the set of achievable points, whose peak delivery latency meets the converse bound.

Without loss of generality, we assume $R_{p1} \geq R_{p2}$. Based on the achievable scheme for problem $\mathcal{Q}^c(M_1,M_2)$, we consider the rate of the message transmitted over the shared common link, $r_c$, for a given $(M_1,M_2,r_{p1},r_{p2})$ tuple.

The seven points considered in Section \ref{Nan73} for the achievability of  problem $\mathcal{Q}^c(M_1,M_2)$, i.e., points A to E, correspond to the following seven points in the format $(M_1,M_2,r_{p1},r_{p2},r_c)$: $P_A=(0,0,0,0,2)$, $P_B=(\frac{N}{2},\frac{N}{2},0,0,\frac{1}{2})$, $P_C=(N,N,0,0,0)$, $P_D=(N,0,0,0,1)$, $P_E=(0,N,0,0,1)$, $P_F=(N-1,0,0,0,1)$ and $P_G=(0,N-1,0,0,1)$. We add five new points:
\begin{enumerate}
\item Point $P_H=(0,0,1,1,0)$. This is the case with no caches at the users. The server transmits $W_{d1}$ to User 1 and $W_{d_2}$ to User 2 via the corresponding private links, respectively.
\item Point $P_I=(0,0,1,0,1)$. In this case the server transmits $W_{d1}$ to User 1 via its private link and $W_{d_2}$ to User 2 via the shared common link.
\item Point $P_J=(0,0,0,1,1)$. This case is symmetric to Point $P_I$.
\item Point $P_K=(0,N,1,0,0)$. This is the case in which User $2$ can cache the entire library, while User $1$ has no cache. The server transmits $W_{d_1}$ to User 1 via its private link.
\item Point $P_L=(N,0,0,1,0)$. This case is symmetric to Point $P_K$.
\end{enumerate}
These twelve points are achievable for problem $\mathcal{Q}(M_1, M_2, R_c, R_{p1}, R_{p2})$.\par
By using memory-sharing for the cache capacity values and time-sharing for the transmitted rates $(r_{p1},r_{p2})$, the convex hull of these twelve points and the corresponding $r_c$ value, i.e., $(M_1,M_2,r_{p1},r_{p2})$ as the independent variables and $r_c$ as the dependent variable, are also achievable. Therefore, we obtain a set of achievable tuples for problem $\mathcal{Q}(M_1, M_2, R_c, R_{p1}, R_{p2})$, denoted by $\Delta$.\par
For a $(M_1,M_2,r_{p1},r_{p2})$ tuple, let $f(M_1, M_2, r_{p1}, r_{p2})$ be the smallest rate $\bar{r}_c$ in $\Delta$, i.e.,
\begin{equation}
  \bar{r}_c=f(M_1, M_2, r_{p1}, r_{p2})=\min\{r_c:(M_1,M_2,r_{p1},r_{p2},r_c)\in \Delta\}.\nonumber
\end{equation}
To obtain $f(M_1, M_2,r_{p1}, r_{p2})$ in closed form, we consider its projection for fixed values of $(r_{p1}, r_{p2})$, and derive $f_{(r_{p1}, r_{p2})}(M_1, M_2)$ in closed form. Before we delve into the details, we provide some insights on the achievable scheme corresponding to $f_{(r_{p1}, r_{p2})}(M_1, M_2)$. Suppose that rates $0 \leq r_{pk} \leq 1$, $k=1,2$, will be transmitted over the private link.

\emph{How to use the private links:} The private links will be used to transmit part of the desired messages in an uncoded fashion. Then the delivery strategy is designed for file sizes reduced by the rates transmitted over the private links. For example, for $r_{p1} \geq r_{p2}$, we split each file into three parts $W_{i}^{c},W_{i}^{p1}$ and $W_{i}^{p12}, i=1,\ldots,N$, with sizes $l_{1} ,l_{2}-l_{1},1-l_{2}$, respectively, where $l_1 \triangleq 1-r_{p1}$ and $l_2 \triangleq1-r_{p2}$. In the delivery phase, the server transmits $\{W_{d_1}^{p1},W_{d_1}^{p12}\}$ and $W_{d_2}^{p12}$ to Users 1 and 2, respectively, via their private links. Thus, we only need to deliver $(W_{d_1}^c, W_{d_2}^c)$ among sub-files $\{W_1^c, W_2^c, \cdots, W_N^c\}$ to Users 1 and 2, and $W_{d_2}^{p1}$ among sub-files $\{W_1^{p1}, W_2^{p1}, \cdots, W_N^{p1}\}$ to User 2 over the shared links.

\emph{How to deal with the sub-files from $\{W_1^{p1}, W_2^{p1}, \cdots, W_N^{p1}\}$ requested by one user only:}  Memory-sharing is performed among certain special achievable points. In each point, the achievable scheme is to either transmit $W_{d_2}^{p1}$ uncoded through the shared common link, or cache all files $\{W_1^{p1}, W_2^{p1}, \cdots, W_N^{p1}\}$ (of file size $l_2-l_1$) in the cache of User 2. The caching and delivery strategy over the common shared link for files $\{W_1^c, W_2^c, \cdots, W_N^c\}$ (of file size $l_1$) is the same as those proposed for  problem $\mathcal{Q}^c(M_1,M_2)$.

We obtain the following lemma for the closed-form expression of $f_{(r_{p1}, r_{p2})}(M_1, M_2)$.
\begin{figure}
\centering
\subfigure[$r_{p1} \geq r_{p2}, N \geq 3$,]
{ \includegraphics[width=2.25in]{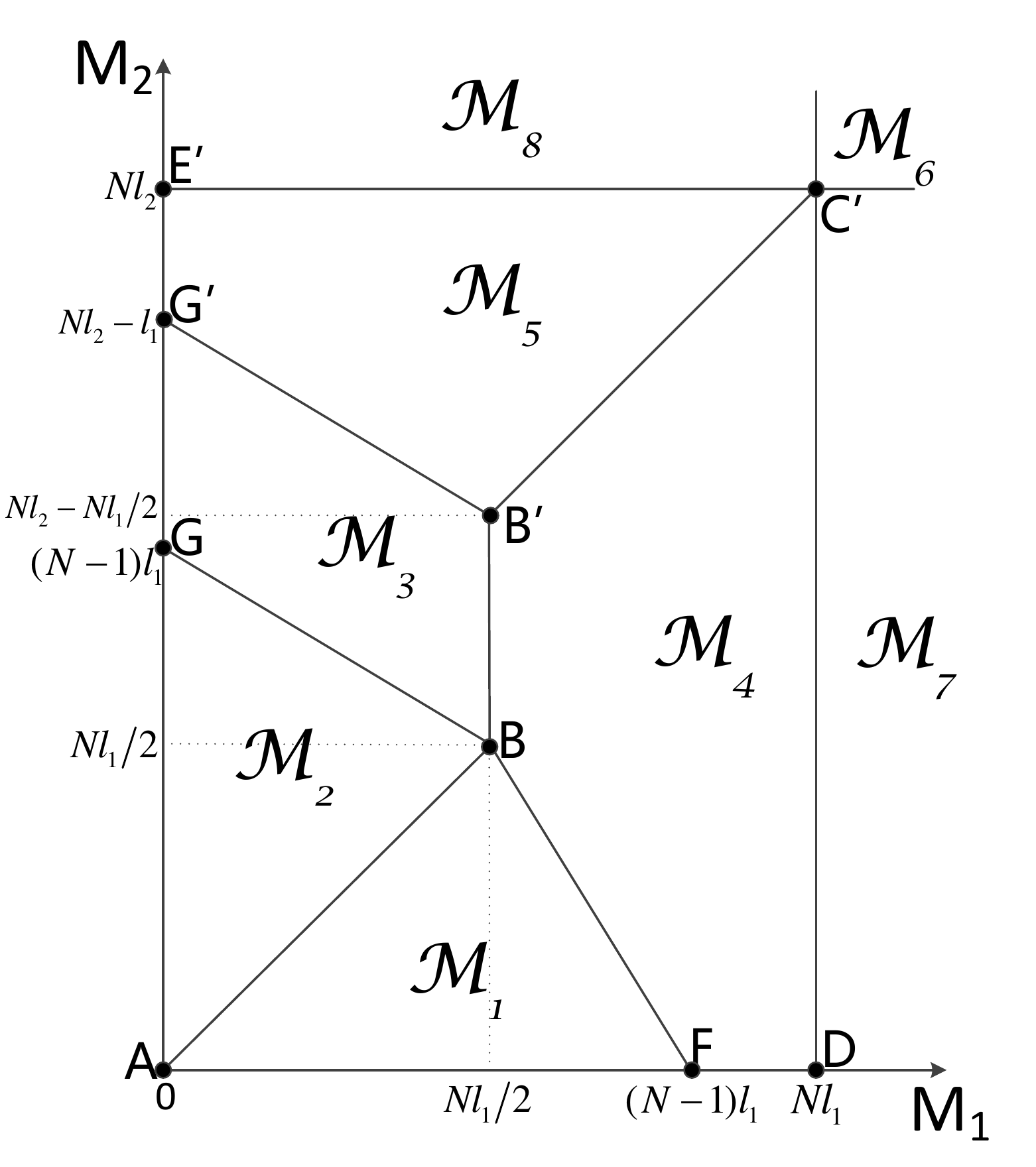}\label{Fig:M_1M_2_case1}}
%\hspace{1in}
\subfigure[$r_{p1} \leq r_{p2}, N \geq 3$,]{
\includegraphics[width=2.25in]{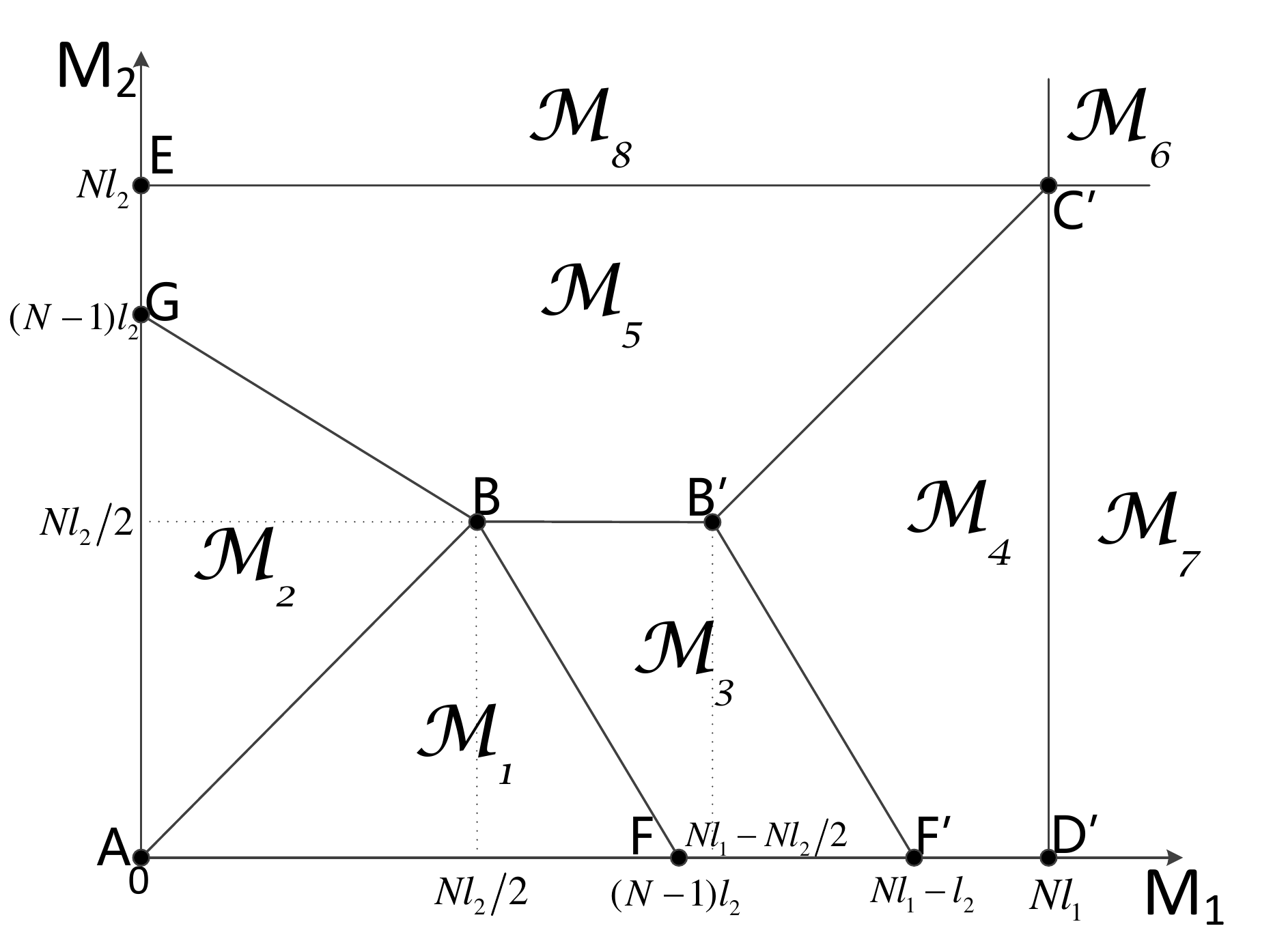}\label{Fig:M_1M_2_case2}}
\caption{The illustration of possible $(M_1,M_2)$ pairs for arbitrary $r_{p1},r_{p2}$ when $N \geq 3$.}
\label{jiemiantuN32}
\end{figure}
\begin{Lemma} \label{Nan120}
For a given $(r_{p1}, r_{p2})$ pair with $r_{p1} \geq r_{p2}$, by memory-sharing among the nine points illustrated in Fig. \ref{Fig:M_1M_2_case1}, the smallest achievable rate over the shared common link, $\bar{r}_{c}=f_{(r_{p1}, r_{p2})}(M_1, M_2)$, is given as
 \begin{align}
\bar{r}_{c} =
\begin{cases}
%l_{1}+l_{2}-\frac{3M_2}{N}-\frac{M_1-M_2}{N-1} =
2-r_{p1}-r_{p2}-\frac{3M_2}{N}-\frac{M_1-M_2}{N-1} & (M_{1},M_{2})\in\mathcal{M}_{1}(r_{p1}, r_{p2})\\
%l_{1}+l_{2}-\frac{3M_1}{N}-\frac{M_2-M_1}{N-1}=
2-r_{p1}-r_{p2}-\frac{3M_1}{N}-\frac{M_2-M_1}{N-1} & (M_{1},M_{2})\in\mathcal{M}_{2}(r_{p1}, r_{p2})\\
%\frac{N-1}{N}l_{1}+l_{2}-\frac{2(N-1)M_{1}}{N^2}-\frac{M_{2}}{N}=
\frac{2N-1}{N}-\frac{N-1}{N}r_{p1}-r_{p2}-\frac{2(N-1)M_{1}}{N^2}-\frac{M_{2}}{N} &(M_{1},M_{2})\in\mathcal{M}_{3}(r_{p1}, r_{p2})\\
%l_{2}-\frac{M_{2}}{N}=
1-r_{p2}-\frac{M_{2}}{N} & (M_{1},M_{2})\in\mathcal{M}_{4}(r_{p1}, r_{p2})\\
%l_{1}-\frac{M_{1}}{N}=
1-r_{p1}-\frac{M_{1}}{N} & (M_{1},M_{2})\in\mathcal{M}_{5}(r_{p1}, r_{p2})
\end{cases},  \label{Nan100}
\end{align}
where the regions $\mathcal{M}_{1}(r_{p1}, r_{p2})$ to $\mathcal{M}_{5}(r_{p1}, r_{p2})$ are shown in Fig \ref{Fig:M_1M_2_case1}.\par

By symmetry, for a given $(r_{p1}, r_{p2})$, where $r_{p1} \leq r_{p2}$, the smallest achievable rate on the shared common link, $\bar{r}_c$,  is given by
 \begin{align}
\bar{r}_{c}=
\begin{cases}
%l_{1}+l_{2}-\frac{3M_2}{N}-\frac{M_1-M_2}{N-1}=
2-r_{p1}-r_{p2}-\frac{3M_2}{N}-\frac{M_1-M_2}{N-1} & (M_{1},M_{2})\in\mathcal{M}_{1}(r_{p1}, r_{p2})\\
%l_{1}+l_{2}-\frac{3M_1}{N}-\frac{M_2-M_1}{N-1}=
2-r_{p1}-r_{p2}-\frac{3M_1}{N}-\frac{M_2-M_1}{N-1} & (M_{1},M_{2})\in\mathcal{M}_{2}(r_{p1}, r_{p2})\\
%\frac{N-1}{N}l_{2}+l_{1}-\frac{2(N-1)M_{2}}{N^2}-\frac{M_{1}}{N}=
\frac{2N-1}{N}-\frac{N-1}{N}r_{p2}-r_{p1}-\frac{2(N-1)M_{2}}{N^2}-\frac{M_{1}}{N} &(M_{1},M_{2})\in\mathcal{M}_{3}(r_{p1}, r_{p2})\\
%l_{2}-\frac{M_{2}}{N}=
1-r_{p2}-\frac{M_{2}}{N} & (M_{1},M_{2})\in\mathcal{M}_{4}(r_{p1}, r_{p2})\\
%l_{1}-\frac{M_{1}}{N}=
1-r_{p1}-\frac{M_{1}}{N} & (M_{1},M_{2})\in\mathcal{M}_{5}(r_{p1}, r_{p2})
%0 & (M_{1},M_{2})\in\mathcal{M}_{6}(r_{p1}, r_{p2})
\end{cases}, \label{Nan101}
\end{align}
where the regions $\mathcal{M}_{1}(r_{p1}, r_{p2})$ to $\mathcal{M}_{5}(r_{p1}, r_{p2})$ are shown in Fig. \ref{Fig:M_1M_2_case2}.
\end{Lemma}

The proof of Lemma \ref{Nan120} is provided in Appendix \ref{Nan121}.
Note that \eqref{Nan100} and \eqref{Nan101} achieve the lower bound of \eqref{hcsgchenbound}, \eqref{hcsgczbound} and the cut-set bound. For an arbitrary $(M_{1},M_{2})$ pair, $0\leq M_{1}\leq N,0\leq M_{2}\leq N$, the set $\Delta$, i.e., the three-dimensional achievable region of $(r_{p1}, r_{p2}, r_c)$, is characterized by (\ref{Nan100}) and (\ref{Nan101}). The remaining task is to find the $(M_1,M_2,r_{p1}, r_{p2}, r_c)$ tuple within the achievable region $\Delta$ that minimizes $T=\max{\{\frac{r_{p1}}{R_{p1}},\frac{r_{p2}}{R_{p2}},\frac{r_{c}}{R_{c}}\}}$.
\begin{Lemma} \label{Nan111}
For any $(M_1, M_2, R_c, R_{p1}, R_{p2})$, there exists an achievable scheme $(M_1, M_2,r_{p1}, r_{p2}, r_c)$ in $\Delta$ with a delivery latency equal to one of the six terms in (\ref{hcsggen}).
\end{Lemma}\par

The proof of Lemma \ref{Nan111} is provided in Appendix \ref{Nan110}.

This completes the achievability part of Theorem \ref{hcsg} for $N \geq 3$ and $R_{p1} \geq R_{p2}$. Before we proceed to the achievability for $N=2$, we make the following connection between the achievability scheme proposed here and the one in \cite{yang2016coded}.

\emph{Remark}: In \cite{yang2016coded} the authors study the caching problem in which the users request different quality descriptions of the files, due to, for example, different processing or display capabilities. For given distortion targets $(D_1, D_2)$, assuming $D_1 \geq D_2$ without loss of generality, the authors suggest using scalable coding \cite{Cover_Equitz} of the files in the library at rates $(r_1, r_2)$, such that the \textit{base layer} of rate $r_1$ allows the first receiver to obtain an average reconstruction distortion of $D_1$, while the base layer together with the \textit{refinement layer} of rate $r_2$ allows an average reconstruction distortion of $D_2$ at the second receiver. This successive coding scheme is known to be rate-distortion optimal for Gaussian sources under squared error distortion.

Once we specify how the private links are used, the $(l_1,l_2)$ parameters in our problem correspond to $(r_1,r_2)$ in the achievable scheme of \cite{yang2016coded}, where $r_1$ corresponds to the number of bits transmitted over the common link, while $r_2-r_1$ to the number of bits transmitted over the private link to the user that request a higher quality description. As such, we may make a comparison of the achievable scheme proposed here and the one in \cite{yang2016coded} for $K=2$ users with $N \geq 3$ files. The scheme in \cite{yang2016coded} is a suboptimal memory-sharing scheme between points $A$, $B$, $B'$, $C'$, $D$, $E'$, ignoring the three points $G$, $G'$  and $F$. We can show that memory-sharing among all the nine points is optimal for the coded caching with heterogeneous distortion requirements problem for $K=2, N \geq 3$, and a converse is provided in Appendix \ref{conversedistortion}. %We state the results formally in the following theorem.

\begin{Theorem}\label{theodistortion}
For the coded caching problem with heterogeneous distortion requirements, defining $l_k=\frac{1}{2} \log \frac{\sigma^2}{D_k}$, $k=1,2$, the optimal cache capacity-delivery trade-off is given by
\begin{align}
  R^*(M_{1},M_{2})=&\max\left\{l_{1}+l_{2}-\frac{3M_2}{N}-\frac{M_1-M_2}{N-1},l_{1}+l_{2}-\frac{3M_1}{N}-\frac{M_2-M_1}{N-1},l_{2}-\frac{M_{2}}{N},l_{1}-\frac{M_{1}}{N},\right.\nonumber\\
  &\left.\frac{N-1}{N}l_{1}+l_{2}-\frac{2(N-1)M_{1}}{N^2}-\frac{M_{2}}{N},\frac{N-1}{N}l_{2}+l_{1}-\frac{2(N-1)M_{2}}{N^2}-\frac{M_{1}}{N}\right\}.\nonumber
\end{align}
\end{Theorem}

\subsection{The achievability of Theorem \ref{hcsg} for $N=2$}
\begin{figure}
\centering
\includegraphics[width=1.6in]{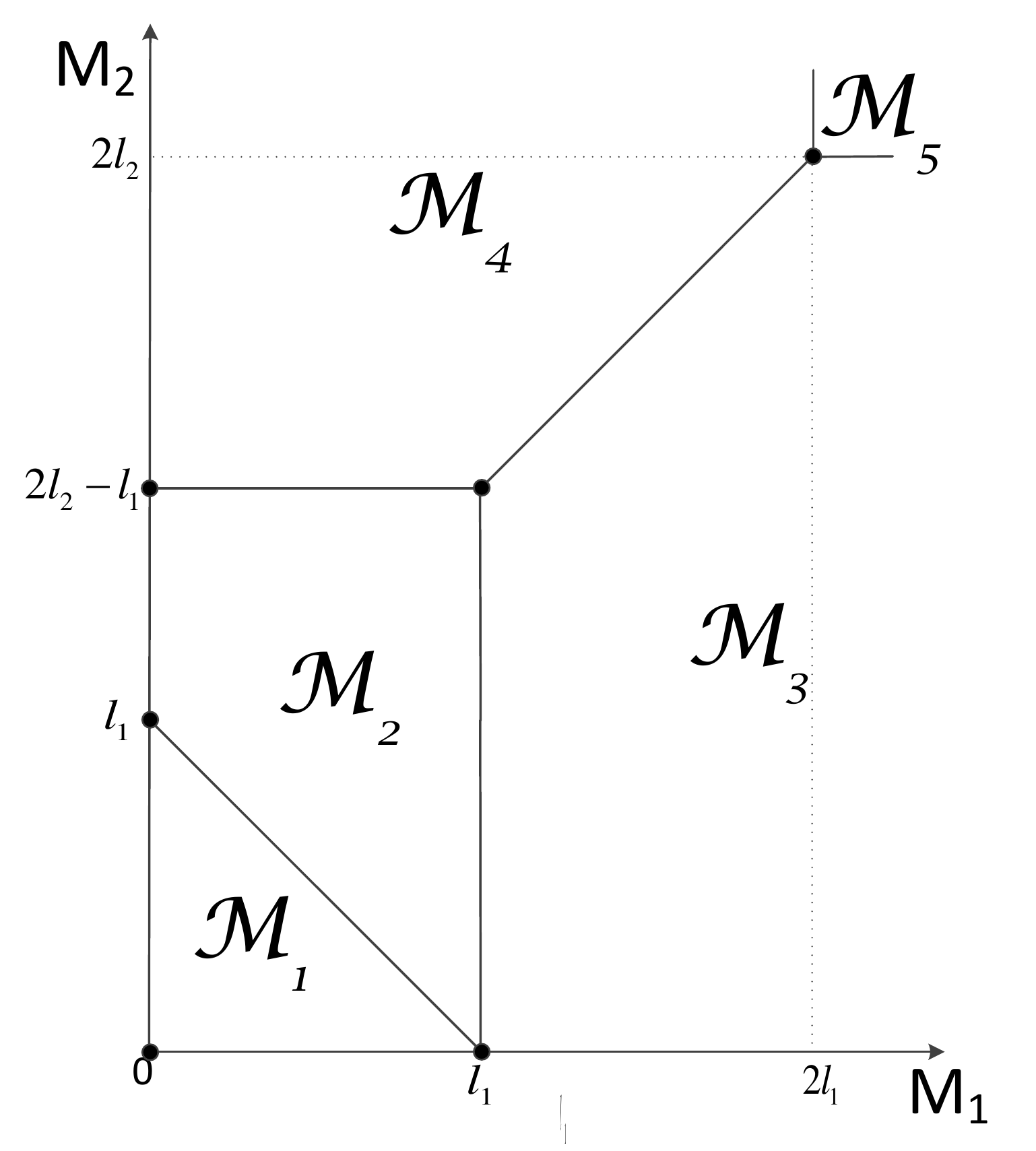}
\caption{The illustration of possible $r_{p1},r_{p2}$ values that satisfy $r_{p1}\geq r_{p2}$ for $N= 2$.}
 \label{jiemiantuN22}
\end{figure}

Based on the above discussion of the similarity between the studied problem  and that of \cite{yang2016coded}, we can use the optimal achievability  found in \cite[Section III.B]{yang2016coded} and obtain the smallest achievable rate on the shared common link, $r_c$, as follows:
\begin{align}
\bar{r}_c=
\begin{cases}
l_{1}+l_{2}-M_{1}-M_2=2-r_{p1}-r_{p2}-M_1-M_2 & (M_{1},M_{2})\in\mathcal{M}_{1}(r_{p1}, r_{p2})\\
\frac{l_{1}}{2}+l_{2}-\frac{M_1}{2}-\frac{M_2}{2}=\frac{3-r_{p1}-2r_{p2}-M_{1}-M_{2}}{2} & (M_{1},M_{2})\in\mathcal{M}_{2}(r_{p1}, r_{p2})\\
l_{2}-\frac{M_{2}}{2}=1-r_{p2}-\frac{M_{2}}{2} & (M_{1},M_{2})\in\mathcal{M}_{3}(r_{p1}, r_{p2})\\
l_{1}-\frac{M_{1}}{2}=1-r_{p1}-\frac{M_{1}}{2} & (M_{1},M_{2})\in\mathcal{M}_{4}(r_{p1}, r_{p2})
\end{cases},\nonumber
\end{align}
where $\mathcal{M}_{1}(r_{p1}, r_{p2})$ to $\mathcal{M}_{4}(r_{p1}, r_{p2})$ are shown in Fig \ref{jiemiantuN22} in the next page.

Similarly to the discussion on the $N \geq 3$ case, we find the achievable $T=\max \left\{\frac{r_{p1}}{R_{p1}},\frac{r_{p2}}{R_{p2}},\frac{r_{c}}{R_{c}}\right\}$ to coincide with (\ref{hcsgtwo}). Thus, the achievability proof of Theorem \ref{hcsg} is complete.

\section{Conclusions}
We have studied the problem of centralized coded caching for two users with different cache capacities, where, in addition to the shared common link, each user also has a private link from the server. We have characterized the optimal caching and delivery strategies for any number of files in the library. In the case of a shared common link only, we have improved upon the known results in the literature by proposing a new achievable scheme for a special $(M_1,M_2)$ pair, and performing memory-sharing among a total of nine special memory pairs. In the case of two private links in addition to the shared common link, we have shown that it is optimal to use all the capacity available over the private links to transmit the file requested by the corresponding user in an uncoded fashion. A connection between the problem of coded caching with a private link to each user considered here and that of coded caching with heterogeneous distortion requirements studied in \cite{yang2016coded} has also been established, which allowed us extending the proposed results to improve the state of the art in the latter problem as well. 

\begin{appendix}

\subsection{Proof of Lemma \ref{chenbound}} \label{Nan01}
We will provide the proof for (\ref{chenboundeq1}), and (\ref{chenboundeq1}) follows by symmetry. For any caching-delivery scheme, we have
\begin{align}
    r_c+M_1 \geq & H(X_c^{1,2})+H(Z_1) \label{Nan03}\\
    \geq& H(Z_1,X_c^{(1,2)})\nonumber\\
    =&H(Z_1,X_c^{(1,2)},W_1)\label{Nan02}\\
    =&H(W_1)+H(Z_1|W_1)+H(X_c^{(1,2)}|Z_1,W_1)\label{Nan19}\\
    \geq& 1+H(Z_1|W_1)+(1-\frac{1}{N-1}[H(Z_1|W_1)+H(Z_2|W_1)])\label{chenpf1}\\
    \geq& 2+\frac{N-2}{N-1}H(Z_{1}|W_{1})-\frac{1}{N-1} H(Z_{2}|W_{1}),\label{chenpf2}
\end{align}
where (\ref{Nan03}) follows from the problem definition in Section \ref{Sys_Mod}, (\ref{Nan02}) follows from the fact that User 1 can decode $W_1$ from $(Z_1, X_c^{(1,2)})$, (\ref{chenpf1}) is from Lemma \ref{skey}.

Similarly, by exchanging the indices of 1 and 2, we have
\begin{align}
  r_c+M_2
  \geq 2+\frac{N-2}{N-1}H(Z_{2}|W_{1})-\frac{1}{N-1} H(Z_{1}|W_{1}).\label{chenpf3}
\end{align}
By cancelling the term $H(Z_{1}|W_{1})$ in (\ref{chenpf2}) and (\ref{chenpf3}), we obtain
\begin{align}
 M_{1}+r_c+(N-2)[r_c+M_{2}] \geq & 2(N-1)+(N-3) H(Z_{2}|W_{1})\nonumber\\ %\label{chenpf2}\\
 \geq & 2(N-1)+\frac{(N-3)(N-1)}{N} H(Z_2),\label{chenpf4}
\end{align}
where (\ref{chenpf4}) follows from Lemma \ref{scachelow}.

Hence, following from (\ref{chenpf4}), we have
\begin{equation}
  NM_{1}+(2N-3)M_{2}+N(N-1)r_c\geq 2N(N-1),\nonumber
\end{equation}
which completes the proof of Lemma \ref{chenbound}.

\subsection{Proof of Lemma \ref{skey}} \label{Nan04}
The proof of Lemma \ref{skey} is given here for completeness, but it follows the proof of \cite[Lemma 1]{tian2018symmetry} very closely. By setting $n=1$  in \cite[Lemma 1]{tian2018symmetry} and not using symmetry, i.e., \cite[Eqn. (13)]{tian2018symmetry}, to replace $Z_2$ with $Z_1$, we would obtain Lemma \ref{skey}. For completeness, the proof of Lemma \ref{skey} is as follows:

In the problem $\mathcal{Q}(M_1, M_2)$, we have
\begin{align}
  (N-1)H(X_c^{(1,2)}|Z_{1},W_1) = &\sum_{i=2 }^N  H(X_c^{(1,i)}|Z_{1},W_1)\label{Nan06}\\
  \geq& H(X_c^{(1,[2:N])}|Z_1,W_1)\nonumber\\
 % \geq& H(X_c^{(1,[2:N])},X_{p1}^{(1,[2:N])},Z_2|Z_1,W_1)-H(Z_2|Z_1,W_1)\nonumber\\
  \geq& H(X_c^{(1,[2:N])},Z_2|W_1)-H(Z_1|W1)-H(Z_2|Z_1,W_1)\nonumber\\
  =&H(X_c^{(1,[2:N])},Z_2,W_{[2:N]}|W_1)-H(Z_2|W1)-H(Z_1|Z_2,W_1)\label{Nan07}\\
  \geq& (N-1)-[H(Z_2|W_1)+H(Z_1|W_1)],\label{Nan08}
\end{align}
where (\ref{Nan06}) is from Lemma \ref{tiansym}, (\ref{Nan07}) follows because given $(X_c^{(1,[2:N])},Z_2)$, User 2 can recover $W_{[2:N]}$, and (\ref{Nan08}) is from $H(X_c^{(1,[2:N])},Z_2|W_{[1:N]})=0$. Thus, we have proved (\ref{Nan12}), and the rest case follows by symmetry.

\subsection{Proof of Lemma \ref{scachelow}} \label{Nan09}

For any $i\in \{1:N-1\}$, we have
\begin{align}
H(W_{[1:i]},Z_{1})-H(W_{[1:i-1]},Z_{1}) = & H(W_i|W_{[1:i-1]},Z_{1}) \nonumber\\
=&H(W_{i+1}|W_{[2:i]},Z_{1})\label{scacheloweqpf1}\\
\geq& H(W_{i+1}|W_{[1:i]},Z_{1})\nonumber\\
=&H(W_{[1:i+1]},Z_{1})-H(W_{[1:i]},Z_{1}),\nonumber
\end{align}
where (\ref{scacheloweqpf1}) is from Lemma \ref{tiansym}.\par
 Then we have
\begin{align}
  \sum_{i=1}^{N-1}(N-i)[H(W_{[1:i]},Z_{1})-H(W_{[1:i-1]},Z_{1})] &\geq \sum_{i=1}^{N-1}(N-i)[H(W_{[1:i+1]},Z_{1})-H(W_{[1:i]},Z_{1})]\nonumber\\
  \Leftrightarrow [\sum_{i=1}^{N-1}H(W_{[1:i]},Z_{1})]-(N-1)H(Z_{1})&\geq [\sum_{i=1}^{N-1}H(W_{[1:i+1]},Z_{1})]-(N-1)H(W_{1},Z_{1})\nonumber\\
  \Leftrightarrow (N-1)H(W_{1},Z_{1})-(N-1)H(Z_{1})&\geq H(W_{[1:N]},Z_{1})-H(W_{1},Z_{1})\nonumber\\
  \Leftrightarrow NH(W_{1},Z_{1})-H(W_{[1:N]})&\geq (N-1)H(Z_{1})\label{scachelowpf2}\\
  \Leftrightarrow NH(Z_{1}|W_{1})&\geq (N-1)H(Z_{1}),\nonumber
\end{align}
where (\ref{scachelowpf2}) is from $H(Z_1|W_{\{1:N\}})=0$. Thus, we have proved (\ref{Nan14}), and the rest case follows from symmetry.

\subsection{Proof of Lemma \ref{czbound}} \label{Nan18}
For User $2$, we have
\begin{align}
   M_2+r_c  \geq& H(Z_2,X_c^{(2,1)})\nonumber\\
 %  =&H(Z_2,X_c^{(2,1)},W_1)\nonumber\\
   =&H(W_1)+H(Z_2|W_1)+H(X_c^{(2,1)}|Z_2,W_1)\label{Nan20}\\
   \geq& 3+\frac{N-2}{N-1}\left[H(Z_{2}|W_{1})+H(Z_{1}|W_{1})\right]-r_c-r_{p1}-M_1,\label{czpf1}
\end{align}
where (\ref{Nan20}) follows from the same steps as (\ref{Nan19}), and (\ref{czpf1}) is from (\ref{skeycoreq2}) in Lemma \ref{skeycor}.\par
And similarly to (\ref{czpf1}), we have
\begin{align}
  M_1+r_c+r_{p1} & \geq H(Z_1,X_c^{(1,2)},X_{p1}^{(1,2)})\nonumber\\
  &=H(W_1)+H(Z_1|W_1)+H(X_c^{(1,2)},X_{p1}^{(1,2)}|Z_1,W_1) \nonumber\\
  &\geq 2+\frac{N-2}{N-1}H(Z_{1}|W_{1})-\frac{1}{N-1} H(Z_{2}|W_{1}),\label{czpf2}
\end{align}
where (\ref{czpf2}) follows from (\ref{skeycoreq1}) in Lemma \ref{skeycor}.

Therefore, by cancelling the term $H(Z_{2}|W_{1})$ in (\ref{czpf1}) and (\ref{czpf2}), we obtain (\ref{hcspczbound}), which completed the proof.

\subsection{The proof of Lemma \ref{skeycor}} \label{Nan22}

In problem $\mathcal{Q}(M_1, M_2, R_c, R_{p1}, 0)$, substituting $X_c^{(i,j)}$ in the proof of (\ref{Nan08}) with $(X_c^{(i,j)},X_{p1}^{(i,j)})$, we get (\ref{skeycoreq1}). Similarly, for (\ref{skeycoreq2}), we have
\begin{align}
  &(N-1)H(X_c^{(2,1)}|Z_{2},W_1)\nonumber\\ 
  =&\sum_{i=2}^N  H(X_c^{(i,1)}|Z_{2},W_1)\label{skeyeqpf3}\\
  \geq& H(X_c^{([2:N],1)}|Z_2,W_1)\nonumber\\
 % \geq& H(X_c^{([2:N],1)},X_{p1}^{([2:N],1)},Z_1|Z_2,W_1)-H(X_{p1}^{([2:N],1)},Z_1|Z_2,W_1)\\
  \geq& H(X_c^{([2:N],1)},X_{p1}^{([2:N],1)},Z_1|W_1)-H(Z_2|W_1)-H(X_{p1}^{([2:N],1)},Z_1|Z_2,W_1)\nonumber\\
  =&H(X_c^{([2:N],1)},X_{p1}^{([2:N],1)},Z_1,W_{[2:N]}|W_1)-H(Z_2|W_1)-H(Z_1|Z_2,W_1)-H(X_{p1}^{([2:N],1)}|Z_1,Z_2,W_1)\nonumber\\
  \geq& (N-1)-[H(Z_2|W_1)+H(Z_1|W_1)]-(N-1)H(X_{p1}^{(2,1)}|Z_1,Z_2,W_1),\label{skeyeqpf4}
\end{align}
where (\ref{skeyeqpf3}) follows from Lemma \ref{tiansym}, and (\ref{skeyeqpf4}) from $H(X_c^{([2:N],1)},X_{p1}^{([2:N],1)},Z_1|W_{[1:N]})=0$ and Lemma \ref{tiansym}.

Finally, we upper bound $H(X_{p1}^{(2,1)}|Z_1,Z_2,W_1)$ as follows:
\begin{align}
  H(X_{p1}^{(2,1)}|& W_{1},Z_{1},Z_{2})  \leq H(X_{p1}^{(2,1)},X_c^{(2,1)}|W_{1},Z_{1},Z_{2}) \nonumber \\
  =& H(X_{p1}^{(2,1)},X_c^{(2,1)},Z_{1},Z_{2},W_1)-H(W_{1},Z_{1},Z_{2}) \nonumber \\
  =& H(X_{p1}^{(2,1)},X_c^{(2,1)},Z_{1},Z_{2},W_2)-H(W_{1},Z_{1},Z_{2}) \nonumber \\
  =& H(X_{p1}^{(2,1)}|W_2,X_c^{(2,1)},Z_{1},Z_{2})+H(W_2,X_c^{(2,1)},Z_{1},Z_{2})-H(W_{2},Z_{1},Z_{2})\label{coneq3}\\
  =& H(X_{p1}^{(2,1)}|W_2,X_c^{(2,1)},Z_{1},Z_{2})+H(X_c^{(2,1)}|W_2,Z_{1},Z_{2})\nonumber \\
  \leq& H(X_{p1}^{(2,1)}|W_2,X_c^{(2,1)},Z_{1})+H(X_c^{(2,1)}|W_2,Z_{1})\nonumber \\
  =& H(X_{p1}^{(2,1)},X_c^{(2,1)}|W_2,Z_{1})\nonumber \\
  =& H(X_{p1}^{(2,1)},X_c^{(2,1)}|Z_{1})-H(W_2)-H(Z_1|W_2)+H(Z_1)\nonumber \\
  \leq& r_c+r_{p1}+M_1-1-H(Z_1|W_1),\label{coneq4}
\end{align}
where (\ref{coneq3}) and (\ref{coneq4}) follow from Lemma \ref{tiansym}.
From (\ref{skeyeqpf4}) and (\ref{coneq4}), we obtain (\ref{skeycoreq2}), which completes the proof.

\subsection{Proof of Lemma \ref{Nan120}} \label{Nan121}
We will characterize $f_{(r_{p1}, r_{p2})}(M_1, M_2)$ for a given $(r_{p1}, r_{p2})$ pair. To do so, we consider the $(M_1,M_2)$ plane for a fixed $(r_{p1}, r_{p2})$ pair, as illustrated in Fig. \ref{jiemiantuN32}(a). The achievability follows from performing memory-sharing among the nine points specified below. These correspond to points $A$ to $G$ in Fig. \ref{fenqu}, plus either transmitting $W_{d_2}^{p1}$ uncoded through the shared common link, or caching all files $\{W_1^{p1}, W_2^{p1}, \cdots, W_N^{p1}\}$ at User 2, which is also reflected in the notation used to refer to these points. Recall that all these points can be achieved from the twelve points $P_A$ to $P_L$ described in Section \ref{sec:general_achieve} via memory-sharing. The points used in memory-sharing and the corresponding fractions for these nine points are given as follows.
\begin{enumerate}
\item Point $A$: it can be achieved by memory-sharing among Points $P_A,P_H$ and $P_I$ with fractions $l_{1},1-l_{2}$ and $l_{2}-l_{1}$, respectively.
\item Point $B$: it can be achieved by memory-sharing among Points $P_B,P_H$ and $P_I$ with fractions $l_{1},1-l_{2}$ and $l_{2}-l_{1}$, respectively.
\item Point $B'$: it can be achieved by memory-sharing among Points $P_B,P_H$ and $P_K$ with fractions $l_{1},1-l_{2}$ and $l_{2}-l_{1}$, respectively.
\item Point $C'$: it can be achieved by memory-sharing among Points $P_C,P_H$ and $P_K$ with fractions $l_{1},1-l_{2}$ and $l_{2}-l_{1}$, respectively.
\item Point $D$: it can be achieved by memory-sharing among Points $P_D,P_H$ and $P_I$ with fractions $l_{1},1-l_{2}$ and $l_{2}-l_{1}$, respectively.
\item Point $E'$: it can be achieved by memory-sharing among Points $P_E,P_H$ and $P_K$ with fractions $l_{1},1-l_{2}$ and $l_{2}-l_{1}$, respectively.
\item Point $F$: it can be achieved by memory-sharing among Points $P_F,P_H$ and $P_I$ with fractions $l_{1},1-l_{2}$ and $l_{2}-l_{1}$, respectively.
\item Point $G$: it can be achieved by memory-sharing among Points $P_G,P_H$ and $P_I$ with fractions $l_{1},1-l_{2}$ and $l_{2}-l_{1}$, respectively.
\item Point $G'$: it can be achieved by memory-sharing among Points $P_G,P_H$ and $P_K$ with fractions $l_{1},1-l_{2}$ and $l_{2}-l_{1}$, respectively.
\end{enumerate}

Next, we present the coding scheme for Points $B$ and $B'$ to illustrate our observation that the schemes either transmit $W_{d_2}^{p1}$ uncoded over the shared common link, or cache all the files $\{W_1^{p1}, W_2^{p1}, \cdots, W_N^{p1}\}$ at User 2. Similarly for the other points.

For point $B$ with $(M_1,M_2,r_c)=(\frac{N}{2}l_{1}, \frac{N}{2}l_{1}, l_{2}-\frac{l_1}{2})$, we use the scheme for Point $B$ of Fig. \ref{fenqu} for subfiles $\{W_i^c, i\in [N]\}$, and transmit $W_{d_2}^{p1}$ through the common link. In other words, each subfile $W_{i}^{c}$ is split into two parts of equal size $(W_i^{c1}, W_i^{c2})$, $i\in [N]$. User $k$ caches $\{W_i^{ck}, i\in [N]\}$, $k=1,2$. In the delivery phase, $\{W_{d_1}^{c2} \oplus W_{d_2}^{c1},W_{d_2}^{p1}\}$ is transmitted over the shared link. 

For point $B'$ with $(M_1, M_2, r_c)=(\frac{N}{2}l_{1}, N l_2-\frac{N}{2}l_{1}, \frac{l_1}{2})$, we also use the scheme for Point $B$ of Fig. \ref{fenqu} for subfiles $\{W_i^c, i\in [N]\}$, i.e., each subfile $W_{i}^{c}$ is split into two parts of equal size $(W_i^{c1}, W_i^{c2})$, $i\in [N]$. Compared with point $B$, instead of transmitting $W_{d_2}^{p1}$ through the common link, we cache $\{W_i^{p1}, i\in [N]\}$ at User 2. In other word, User $k$ caches $\{W_i^{ck}, i\in [N]\}$, $k=1,2$, and furthermore, User 2 caches $\{W_{1}^{p1},W_{2}^{p1},\cdots,W_{N}^{p1}\}$.In the delivery phase, $\{W_{d_1}^{c2} \oplus W_{d_2}^{c1}\}$ is transmitted over the shared link.

In Fig. \ref{jiemiantuN32} (a), for $(M_1, M_2) \in \mathcal{M}_1$, we perform memory-sharing among Points $A$, $B$, $F$; for $(M_1, M_2) \in \mathcal{M}_2$, among Points $A$, $B$ and $G$; for $(M_1, M_2) \in \mathcal{M}_3$, among $B$, $B'$, $G$, $G'$; for $(M_1, M_2) \in \mathcal{M}_4$, among $B$, $B'$, $F$, $D$, $C'$; for $(M_1, M_2) \in \mathcal{M}_5$, among Points $C'$, $B'$, $G'$, $E'$. When $(M_1, M_2) \in \mathcal{M}_6$, the caches at both users are large enough, so we do not need to transmit any data over the shared link. When $(M_1, M_2) \in \mathcal{M}_7$, we waste the extra cache at User 1 and achieve the  same performance as point $(Nl_1, M_2) \in \mathcal{M}_4$. Similarly, when $(M_1, M_2) \in \mathcal{M}_8$, we waste the extra cache at User 2 and achieve the  same performance as point $(M_1, N l_2) \in \mathcal{M}_5$. Hence, we focus on the non-trivial cases of $\mathcal{M}_1 \bigcup \mathcal{M}_2 \bigcup \cdots \bigcup \mathcal{M}_5$, and the memory-sharing expressions are given by (\ref{Nan100}). By symmetry, we can also obtain (\ref{Nan101}).
\par

\subsection{Proof of Lemma \ref{Nan111}} \label{Nan110}
\begin{figure}
\centering
\subfigure[$M_{1}\geq M_{2}$ and $0\leq\frac{R_{p2}}{R_{p1}}\leq1$ ]{ \label{twolineseg-a}
\includegraphics[width=2in]{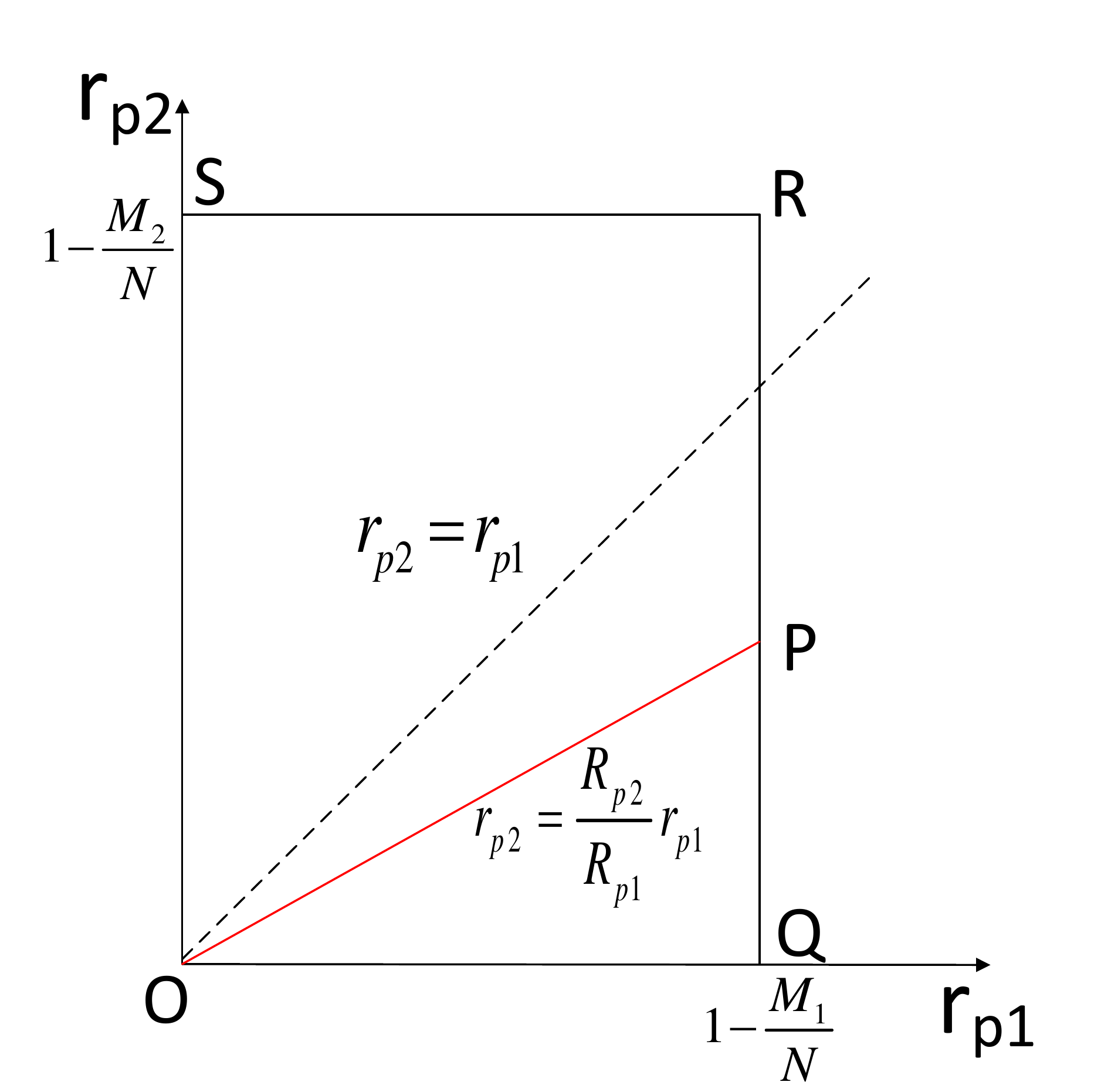}}
%\hspace{1in}
\subfigure[$M_{1}\leq M_{2}$ and $0\leq\frac{R_{p2}}{R_{p1}}\leq\frac{N-M_{2}}{N-M_{1}}$]{ \label{twolineseg-b}
\includegraphics[width=2in]{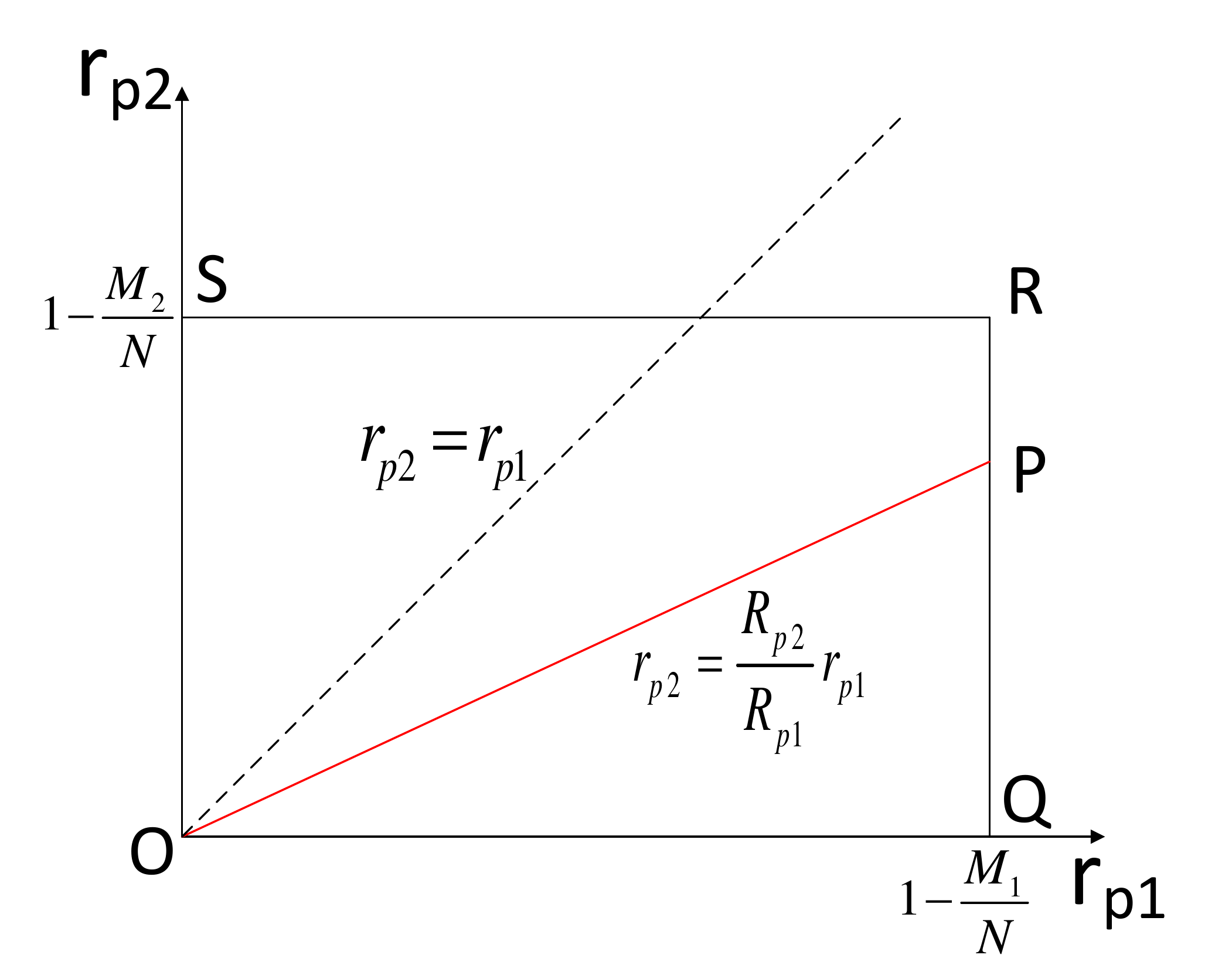}}
\subfigure[$M_{1}\leq M_{2}$ and $\frac{N-M_{2}}{N-M_{1}}\leq\frac{R_{p2}}{R_{p1}}\leq1$]{ \label{twolineseg-c}
\includegraphics[width=2in]{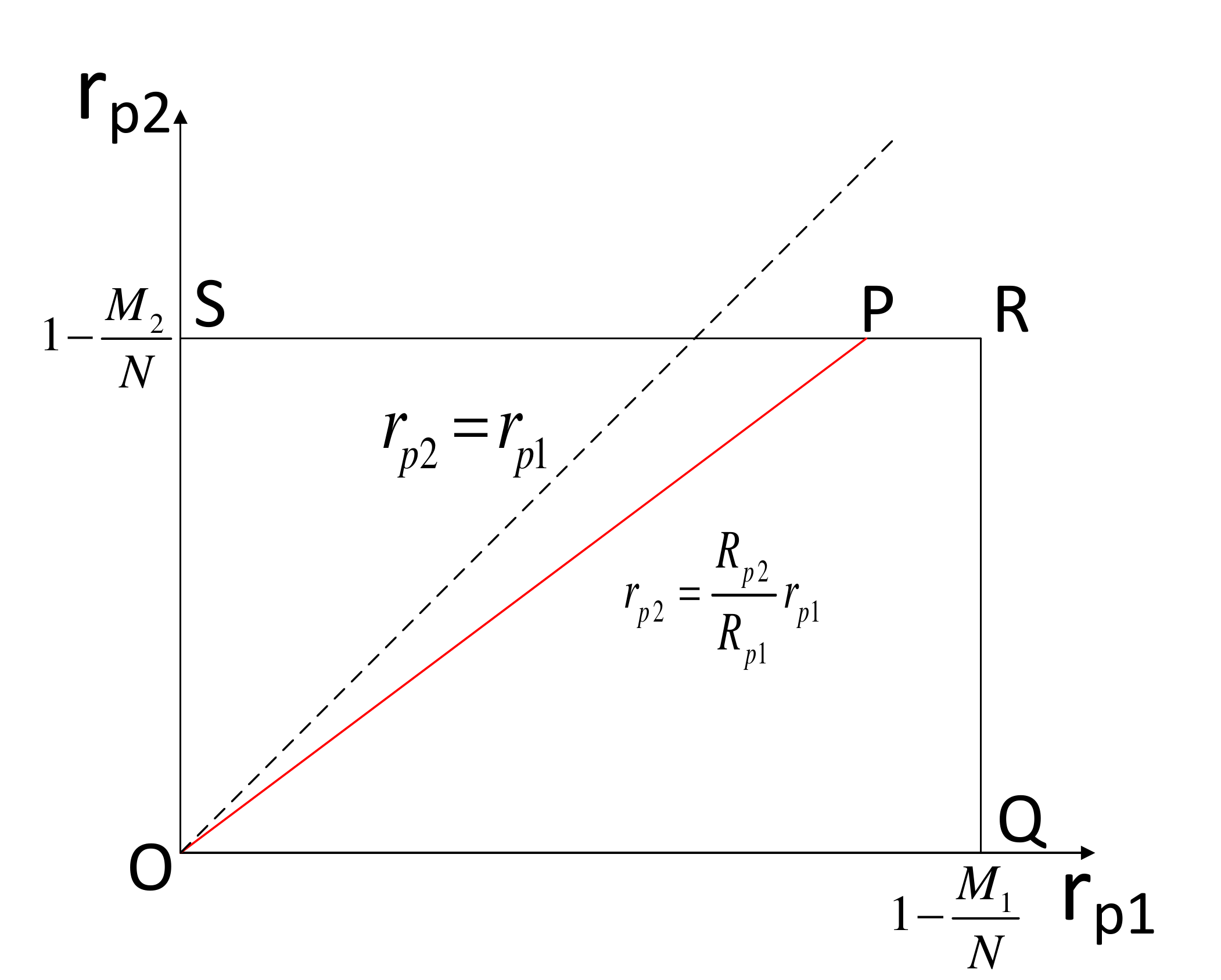}}
\caption{For a fixed $(M_1, M_2)$ pair, the achievable $(r_{p1}, r_{p2})$ region.}
\label{ZhangDeyao}
\end{figure}
In this proof, we consider another projection of $f(M_1,M_2, r_{p1}, r_{p2})$ where we fix the pair $(M_1, M_2)$ and focus on the function $f_{(M_1, M_2)}(r_{p1}, r_{p2})$ for the remaining parameters $(r_{p1}, r_{p2})$.

Note that $\bar{r}_c=f_{(M_1, M_2)}(r_{p1}, r_{p2})$ can be found explicitly from (\ref{Nan100}) or (\ref{Nan101}), albeit the expressions may be tedious to write explicitly. However, we do not need the explicit expression of $f_{(M_1,M_2)}(\cdot)$, only its following properties: i) Since $f(M_1, M_2,r_{p1}, r_{p2})$ is continuous and the closed-form expression of $f_{(r_{p1}, r_{p2})}(M_1,M_2)$ in \eqref{Nan100} and  \eqref{Nan101} is monotonically decreasing in $(r_{p1}, r_{p2})$, $f_{(M_1, M_2)}(r_{p1}, r_{p2})$ is a continuous and monotonically decreasing function of $(r_{p1}, r_{p2})$, where the monotonicity is defined as $f_{(M_1,M_2)}(r_{p1}, r_{p2}) \geq f_{(M_1,M_2)}(r'_{p1}, r'_{p2})$ if $r_{p1} \leq r'_{p1}, r_{p2} \leq r'_{p2}$; ii) The value of $f_{(M_1,M_2)}(r_{p1}, r_{p2})$ can take only one of the five values in (\ref{Nan100}) or (\ref{Nan101}).

For a given and fixed $(M_1, M_2)$ pair, we pick an achievable $(r_{p1}, r_{p2}, r_c)$ tuple as follows:

Note that, since none of the points with coded cache, i.e., $P_F$ and $P_G$, lie on the boundary in this projection, it is sufficient to only consider the rectangle $0\leq r_{pi} \leq 1-\frac{M_i}{N}$, $i=1,2$, since the rate $r_{pi}=1-\frac{M_i}{N}$, $i=1,2$, is enough for User $i$, $i=1,2$, to recover the file, respectively.

We have the following cases as shown in Figure \ref{ZhangDeyao}, which shows the projection to the space with parameters $(r_{p1}, r_{p2})$:
\begin{itemize}
\item \textbf{Case 1:} $\{M_1>M_2, 0\leq \frac{R_{p2}}{R_{p1}}\leq 1\}$ or $\{M_1<M_2,\frac{R_{p2}}{R_{p1}} \leq \frac{N-M_2}{N-M_1}\}$, i.e., Fig. \ref{ZhangDeyao}(a) and (b). For this case, we further have the following two sub-cases:
\begin{itemize}
\item
$
0\leq \frac{R_{p1}}{R_{c}+R_{p2}}\leq\frac{N-M_{1}}{N-M_{2}}
$:
The achievable $(r_{p1}, r_{p2})$ we pick is inside the rectangle, and also lies on line $r_{p2}=\frac{R_{p2}}{R_{p1}} r_{p1}$, i.e., it is the line segment of $OP$ in Fig. \ref{ZhangDeyao}(a) or Fig. \ref{ZhangDeyao}(b).

Consider the following function of $r_{p1}$:
\begin{align}
g_1(r_{p1}) \triangleq \frac{r_{p1}}{f_{(M_1,M_2)}(r_{p1}, \frac{R_{p2}}{R_{p1}} r_{p1})+\frac{R_{p2}}{R_{p1}} r_{p1}}.\nonumber
\end{align}
Since $f_{(M_1,M_2)}(r_{p1}, r_{p2})$ is continuous and monotonically decreasing, $g_1(r_{p1})$ is continuous and montonically increasing.
At the point $O$ in Fig. \ref{ZhangDeyao}(a) and \ref{ZhangDeyao}(b), i.e.,\\ $(r_{p1}, r_{p2})=(0,0)$, $g_1(0)=0$. At the point $P$ in Fig. \ref{ZhangDeyao}(a) and \ref{ZhangDeyao}(b), i.e., $(r_{p1}, r_{p2})=\left(1-\frac{M_1}{N},\frac{R_{p2}}{R_{p1}}\left(1-\frac{M_1}{N} \right) \right)$, we have $M_1=N(1-r_{p1})=Nl_1$, which in region $\mathcal{M}_4(r_{p1}, r_{p2})$ in \eqref{Nan100}. This gives us $f_{(M_1,M_2)}(r_{p1}, \frac{R_{p2}}{R_{p1}} r_{p1})+\frac{R_{p2}}{R_{p1}} r_{p1}=1-\frac{M_2}{N}$, and as a result, $g_1(1-\frac{M_1}{N})=\frac{N-M_1}{N-M_2}$. Since we are considering the case $0\leq \frac{R_{p1}}{R_{c}+R_{p2}}\leq\frac{N-M_{1}}{N-M_{2}}
$,
we may find a $\tilde{r}_{p1}$, where $\left(\tilde{r}_{p1}, \frac{R_{p2}}{R_{p1}} \tilde{r}_{p1}\right)$ lies on the line segment $OP$ in Fig. \ref{ZhangDeyao}(a) and \ref{ZhangDeyao}(b), that satisfies
\begin{align}\label{newcdm1}
g_1(\tilde{r}_{p1})=\frac{R_{p1}}{R_{c}+R_{p2}},\nonumber
\end{align}
and the $(r_{p1}, r_{p2}, r_c)$ point we pick to calculate $T=\max{\{\frac{r_{p1}}{R_{p1}},\frac{r_{p2}}{R_{p2}},\frac{r_{c}}{R_{c}}\}}$ is $(\hat{r}_{p1}, \hat{r}_{p2}, \hat{r}_c)=\left(\tilde{r}_{p1}, \frac{R_{p2}}{R_{p1}} \tilde{r}_{p1}, f_{(M_1,M_2)} (\tilde{r}_{p1}, \frac{R_{p2}}{R_{p1}} \tilde{r}_{p1}) \right)$. Note that this point satisfies
\begin{align}
\frac{\hat{r}_{p1}}{R_{p1}}=\frac{\hat{r}_{p2}}{R_{p2}}=\frac{\hat{r}_{c}}{R_{c}}.
\end{align}
Since $\left(\tilde{r}_{p1}, \frac{R_{p2}}{R_{p1}} \tilde{r}_{p1}\right)$ can take all values on the line segment $OP$ for some $(R_{p1},R_{p2},R_c)$, then the pair $(M_1,M_2)$ can appear in these five regions $\mathcal{M}_1,\cdots,\mathcal{M}_5$ in (\ref{Nan100})  for some $(R_{p1},R_{p2},R_c)$. Therefore, since the value of $f_{M_1,M_2}(\hat{r}_{p1}, \hat{r}_{p2})$ can take only one of the five corresponding values in (\ref{Nan100}), combining with \eqref{newcdm1}, we see that $T=\max{\left\{\frac{\hat{r}_{p1}}{R_{p1}},\frac{\hat{r}_{p2}}{R_{p2}},\frac{\hat{r}_{c}}{R_{c}}\right\}}$ can only take one of the following values
\begin{align}
&\left\{\frac{2-\frac{3M_2}{N}-\frac{M_1-M_2}{N-1}}{R_c+R_{p1}+R_{p2}}, \frac{2-\frac{3M_1}{N}-\frac{M_2-M_1}{N-1}}{R_c+R_{p1}+R_{p2}},
\frac{N(2N-1)-2(N-1)M_{1}-NM_{2}}{N^{2}(R_c+R_{p2})+N(N-1)R_{p1}}, \right.\nonumber\\
& \left.\frac{1-\frac{M_1}{N}}{R_c+R_{p1}}, \frac{1-\frac{M_2}{N}}{R_c+R_{p2}}\right\}. \label{Nan102}
\end{align}
Note that, in this sub-case, it is easy to check that the optimal latency $T^*$ showed in \eqref{hcsggen} is equal to the maximum value of \eqref{Nan102}. Therefore, we have shown that $T=T^*$ in this sub-case due to the fact that $T^*$ is the lower bound of $T$.
%%%%%%%%%%%%%
\item $\frac{R_{p1}}{R_{c}+R_{p2}} >\frac{N-M_{1}}{N-M_{2}}
$: The achievable $(r_{p1}, r_{p2})$ lies on the line segment $QR$ in Fig. \ref{ZhangDeyao}(a) or Fig. \ref{ZhangDeyao}(b), i.e., $r_{p1}=1-\frac{M_1}{N}$. Now, we pick $r_c$ within the three-dimensional achievable region, and this will determine to which point on line segment $QR$ it corresponds.

Consider the following function of $r_{p2}$:
\begin{align}
g_2(r_{p2}) \triangleq \frac{f_{(M_1,M_2)}(1-\frac{M_1}{N}, r_{p2})}{r_{p2}}.\nonumber
\end{align}
Since $f_{(M_1,M_2)}(r_{p1}, r_{p2})$ is continuous and monotonically decreasing, so is $g_2(r_{p2})$. At point $Q$ in Fig. \ref{ZhangDeyao}(a) and \ref{ZhangDeyao}(b), i.e., $(r_{p1}, r_{p2})=(1-\frac{M_1}{N},0)$, $g_2(0)=\infty$, we have $M_1=N(1-r_{p1})=Nl_1$. At the point $R$ in Fig. \ref{ZhangDeyao}(a) and \ref{ZhangDeyao}(b), i.e., $(r_{p1}, r_{p2})=\left(1-\frac{M_1}{N},1-\frac{M_2}{N} \right)$, we have $M_1=N(1-r_{p1}), M_2=N(1-r_{p2})$ which is the point $C'$ in Fig. \ref{jiemiantuN32}(a) or \ref{jiemiantuN32}((b). This gives us $f_{(M_1,M_2)}\left(1-\frac{M_1}{N},1-\frac{M_2}{N} \right)=0$, and as a result, $g_2(1-\frac{M_2}{N} )=0$. Hence, we may find a point $(1-\frac{M_1}{N}, \tilde{r}_{p2})$ on line segment $QR$ that satisfies
\begin{align}
g_2(\tilde{r}_{p2})=\frac{R_c}{R_{p2}},\nonumber
\end{align}
and the $(r_{p1}, r_{p2}, r_c)$ point we pick to calculate $T=\max{\{\frac{r_{p1}}{R_{p1}},\frac{r_{p2}}{R_{p2}},\frac{r_{c}}{R_{c}}\}}$ is $(\hat{r}_{p1}, \hat{r}_{p2}, \hat{r}_c)=\left(1-\frac{M_1}{N}, \tilde{r}_{p2}, f_{(M_1,M_2)} (1-\frac{M_1}{N}, \tilde{r}_{p2}) \right)$. Note that this point satisfies
\begin{align}\label{newcdm2}
\frac{\hat{r}_{p2}}{R_{p2}}=\frac{\hat{r}_{c}}{R_{c}} \geq \frac{\hat{r}_{p1}}{R_{p1}}.
\end{align}
where the last $\geq$ follows from $\tilde{r}_{p2}+ f_{(M_1,M_2)} (1-\frac{M_1}{N}, \tilde{r}_{p2})=1-\frac{M_2}{N}$ and $\frac{R_{p1}}{R_{c}+R_{p2}} >\frac{N-M_{1}}{N-M_{2}}$.\par
In this sub-case, $(M_1,M_2)$ is always in the line segment of  $C'D$ in Fig. \ref{jiemiantuN32} (a) or $C'D'$ in Fig. \ref{jiemiantuN32} (b), i.e., $M_1=N(1-r_{p1})=Nl_1$. Therefore, the value of $f_{(M_1,M_2)} (1-\frac{M_1}{N}, \tilde{r}_{p2})$ is $1-\tilde{r}_{p2}-\frac{M_2}{N}$. Combining with \eqref{newcdm2},  we see that $T=\max{\left\{\frac{\hat{r}_{p1}}{R_{p1}},\frac{\hat{r}_{p2}}{R_{p2}},\frac{\hat{r}_{c}}{R_{c}}\right\}}$ can only take the following value
\begin{align}
 T=\frac{1-\frac{M_2}{N}}{R_c+R_{p2}}.\nonumber 
\end{align}
\end{itemize}
Note that, in this sub-case, it is easy to check that the optimal latency $T^*$ showed in \eqref{hcsggen} is equal to $T$.

\item For the remaining case of Fig. \ref{ZhangDeyao} (c), we again have two sub-cases:
\begin{itemize}
\item $0\leq \frac{R_{p2}}{R_{c}+R_{p1}}\leq\frac{N-M_{2}}{N-M_{1}}$:
consider the function of $r_{p1}$,
\begin{align}
g_3(r_{p1}) \triangleq \frac{\frac{R_{p2}}{R_{p1}} r_{p1}}{f_{(M_1,M_2)}(r_{p1}, \frac{R_{p2}}{R_{p1}} r_{p1})+ r_{p1}}.\nonumber
\end{align}
Due to the fact that $f_{(M_1,M_2)}(r_{p1}, \frac{R_{p2}}{R_{p1}} r_{p1})$ is continuous and monotonically decreasing, $g_2(r_{p1})$ is continuous and montonically increasing.
At the point $O$ in Fig. \ref{ZhangDeyao}(c), i.e., $(r_{p1}, r_{p2})=(0,0)$, $g_3(0)=0$, and at the point $P$ in Fig. \ref{ZhangDeyao} (c), i.e., $(r_{p1}, r_{p2})=\left(\frac{R_{p1}}{R_{p2}}\left(1-\frac{M_2}{N} \right) , 1-\frac{M_2}{N},\right)$, $g_3(\frac{R_{p1}}{R_{p2}}\left(1-\frac{M_2}{N} \right))=\frac{N-M_2}{N-M_1}$. Hence, under the case considered, i.e., $
0\leq \frac{R_{p2}}{R_{c}+R_{p1}}\leq\frac{N-M_{2}}{N-M_{1}}
$,
we may find a $\tilde{r}_{p1}$, where $\left(\tilde{r}_{p1}, \frac{R_{p2}}{R_{p1}} \tilde{r}_{p1}\right)$ is on the line segment of $OP$ in Fig. \ref{ZhangDeyao}(c), that satisfies
\begin{align}
g_3(\tilde{r}_{p1})=\frac{R_{p2}}{R_{c}+R_{p1}},\nonumber
\end{align}
and the $(r_{p1}, r_{p2}, r_c)$ point we pick to calculate $T=\max{\{\frac{r_{p1}}{R_{p1}},\frac{r_{p2}}{R_{p2}},\frac{r_{c}}{R_{c}}\}}$ is $(\hat{r}_{p1}, \hat{r}_{p2}, \hat{r}_c)=\left(\tilde{r}_{p1}, \frac{R_{p2}}{R_{p1}} \tilde{r}_{p1}, f_{(M_1,M_2)} (\tilde{r}_{p1}, \frac{R_{p2}}{R_{p1}} \tilde{r}_{p1}) \right)$.

Similar to the previous case, $T=\max{\left\{\frac{\hat{r}_{p1}}{R_{p1}},\frac{\hat{r}_{p2}}{R_{p2}},\frac{\hat{r}_{c}}{R_{c}}\right\}}$ can only take one of the values in (\ref{Nan102}), and it can be show that, in this sub-case, $T$ achieve the converse bound.

\item  $\frac{R_{p2}}{R_{c}+R_{p1}}\geq\frac{N-M_{2}}{N-M_{1}}$: The achievable $(r_{p1}, r_{p2})$ is on the line segment of $SR$ in Fig. \ref{ZhangDeyao}(c), i.e., $r_{p2}=1-\frac{M_2}{N}$. Now, we pick $r_c$ in the three-dimensional achievable region and this will determine which point on the line segment $SR$ lies.

Consider a function of $r_{p1}$,
\begin{align}
g_4(r_{p1}) \triangleq \frac{f_{(M_1,M_2)}( r_{p1},1-\frac{M_2}{N})}{r_{p1}}.\nonumber
\end{align}
Due to the fact that $f_{(M_1,M_2)}(r_{p1}, r_{p2})$ is continuous and monotonically decreasing, $g_4(r_{p1})$ is continuous and montonically decreasing.
At the point $S$ in Fig. \ref{ZhangDeyao}(c), i.e., $(r_{p1}, r_{p2})=(0, 1-\frac{M_1}{N})$, $g_2(0)=\infty$, and at the point $R$ in Fig. \ref{ZhangDeyao}(c), i.e., $(r_{p1}, r_{p2})=\left(1-\frac{M_1}{N},1-\frac{M_2}{N} \right)$, we have $M_1=N(1-r_{p1}), M_2=N(1-r_{p2})$ which is the point $C'$ in Fig. \ref{jiemiantuN32} (a) or (b) . This gives us $f_{(M_1,M_2)}\left(1-\frac{M_1}{N},1-\frac{M_2}{N} \right)=0$, and as a result, $g_4(1-\frac{M_1}{N} )=0$. Hence, we may find a point $( \tilde{r}_{p1}, 1-\frac{M_2}{N})$ on the line segment $SR$ that satisfies
\begin{align}
g_4(\tilde{r}_{p2})=\frac{R_c}{R_{p1}},\nonumber
\end{align}
and the $(r_{p1}, r_{p2}, r_c)$ point we pick to calculate $T=\max{\{\frac{r_{p1}}{R_{p1}},\frac{r_{p2}}{R_{p2}},\frac{r_{c}}{R_{c}}\}}$ is $(\hat{r}_{p1}, \hat{r}_{p2}, \hat{r}_c)=\left(\tilde{r}_{p1}, 1-\frac{M_2}{N}, f_{(M_1,M_2)} (\tilde{r}_{p1}, 1-\frac{M_2}{N}) \right)$. Note that this point satisfies
\begin{align}
\frac{\hat{r}_{p1}}{R_{p1}}=\frac{\hat{r}_{c}}{R_{c}} \geq \frac{\hat{r}_{p2}}{R_{p2}},\nonumber
\end{align}
where the last $\geq$ follows from $\tilde{r}_{p1}+ f_{(M_1,M_2)} (\tilde{r}_{p1}, 1-\frac{M_2}{N})=1-\frac{M_1}{N}$ and $\frac{R_{p2}}{R_{c}+R_{p1}}\geq\frac{N-M_{2}}{N-M_{1}}$.\par
Since the value of $f_{(M_1,M_2)} (\tilde{r}_{p1}, 1-\frac{M_2}{N})$ is $1-\tilde{r}_{p1}-\frac{M_1}{N}$, similar to the previous case, we see that $T=\max{\left\{\frac{\hat{r}_{p1}}{R_{p1}},\frac{\hat{r}_{p2}}{R_{p2}},\frac{\hat{r}_{c}}{R_{c}}\right\}}$ can only take the following value
\begin{align}
 \frac{1-\frac{M_1}{N}}{R_c+R_{p1}},\nonumber 
\end{align}
which also achieve the converse bound.
\end{itemize}
\end{itemize}

\subsection{Converse proof of Theorem \ref{theodistortion}}\label{conversedistortion}
Firstly, we denote $S_{i}$ as the $i$-th source and $\hat{S}_{i}^k$ as the $i$-th source recovered by the $k$-th user, in which $i=1,\cdots,N$ and $k=1,2$.
Due to the independence of the sources and the constraints of users' decoding, the Lemmas \ref{tiansym} and \ref{scachelow} apply to this model, i.e., there must be an optimal source-index-symmetric caching and delivery code, for which we have:
\begin{align}\label{dtcv1}
  NH(Z_{1}|S_{1})&\geq (N-1)H(Z_{1}),\\
  NH(Z_{2}|S_{1})&\geq (N-1)H(Z_{2}).
\end{align}
Then, similarly to Lemma \ref{skey}, we have
\begin{align}
  &(N-1)H(X_c^{(1,2)}|Z_{1},S_1)  \nonumber\\*
  =&\sum_{i=2 }^N  H(X_c^{(1,i)}|Z_{1},S_1)\label{dtcv2-1}\\
  \geq& H(X_c^{(1,[2:N])}|Z_1,S_1)\nonumber\\
 % \geq& H(X_c^{(1,[2:N])},X_{p1}^{(1,[2:N])},Z_2|Z_1,W_1)-H(Z_2|Z_1,W_1)\nonumber\\
  \geq& H(X_c^{(1,[2:N])},Z_2|S_1)-H(Z_1|S_1)-H(Z_2|Z_1,S_1)\nonumber\\
  =&H(X_c^{(1,[2:N])},Z_2,\hat{S}_{[2:N]}^2|S_1)-H(Z_2|S_1)-H(Z_1|Z_2,S_1)\label{dtcv2-2}\\
  =&H(\hat{S}_{[2:N]}^2|S_1)+H(X_c^{(1,[2:N])},Z_2|\hat{S}_{[2:N]}^2,S_1)-H(Z_2|S_1)-H(Z_1|Z_2,S_1)\nonumber\\
  \geq&H(\hat{S}_{[2:N]}^2|S_1)+H(X_c^{(1,[2:N])},Z_2|S_{[1:N]})-H(Z_2|S_1)-H(Z_1|S_1)\nonumber\\
  \geq&\sum_{i=2}^{N}H(\hat{S}_{i}^2)+H(X_c^{(1,[2:N])},Z_2|S_{[1:N]})-H(Z_2|S_1)-H(Z_1|S_1)\label{dtcv2-3}\\
  \geq& (N-1)l_2-[H(Z_2|S_1)+H(Z_1|S_1)],\label{dtcv2-4}
\end{align}
 where (\ref{dtcv2-1}) follows since we consider source-index-symmetric codes; (\ref{dtcv2-2}) from the recovery of requests from the transmitted messages and cache contents; (\ref{dtcv2-3}) from the independence of sources; and (\ref{dtcv2-4}) from the definition of the rate distortion function.
Similarly,
\begin{equation}
  (N-1)H(X_c^{(2,1)}|Z_{2},S_1)\geq (N-1)l_1-[H(Z_1|S_1)+H(Z_2|S_1)].\nonumber
\end{equation}

Then, similarly to Lemma \ref{skey}, we have
\begin{align}
   r_c+M_1 \geq & H(X_c^{(1,2)})+H(Z_1) \nonumber\\
    \geq& H(Z_1,X_c^{(1,2)})\nonumber\\
    =&H(Z_1,X_c^{(1,2)},\hat{S}_1^1)\nonumber\\
    =&H(\hat{S}_1^1)+H(Z_1|\hat{S}_1^1)+H(X_c^{(1,2)}|Z_1,\hat{S}_1^1)\nonumber\\
    \geq&H(\hat{S}_1^1)+H(Z_1|S_1)+H(X_c^{(1,2)}|Z_1,S_1)\nonumber\\
    \geq& l_1+H(Z_1|S_1)+(l_{2}-\frac{1}{N-1}[H(Z_1|S_1)+H(Z_2|S_1)])\label{dtcv3-1}\\
    \geq& l_1+l_2+\frac{N-2}{N-1}H(Z_{1}|S_{1})-\frac{1}{N-1} H(Z_{2}|S_{1}),\label{dtcv3-2}
\end{align}
where (\ref{dtcv3-1}) follows from (\ref{dtcv2-4}) and the definition of the rate distortion function.

Similarly, by exchanging the indices of 1 and 2, we have
\begin{align}
  r_c+M_2
  \geq l_1+l_2+\frac{N-2}{N-1}H(Z_{2}|S_{1})-\frac{1}{N-1} H(Z_{1}|S_{1}).\label{dtcv3-3}
\end{align}
By cancelling the term $H(Z_{1}|S_{1})$ in (\ref{dtcv3-2}) and (\ref{dtcv3-3}), we obtain for $N\geq 3$
\begin{align}
 &M_{1}+r_c+(N-2)[r_c+M_{2}]\nonumber\\
 \geq& (N-1)(l_1+l_2)+(N-3) H(Z_{2}|S_{1})\nonumber\\
 \geq& (N-1)(l_1+l_2)+\frac{(N-3)(N-1)}{N} H(Z_2),\label{dtcv4-1}
\end{align}
where (\ref{dtcv4-1}) is from (\ref{dtcv1}).

Hence, following from (\ref{dtcv4-1}), we have
\begin{equation}
  NM_{1}+(2N-3)M_{2}+N(N-1)r_c\geq N(N-1)(l_1+l_2).\label{dtcv4-2}
\end{equation}
Symmetrically,
\begin{equation}
  NM_{2}+(2N-3)M_{1}+N(N-1)r_c\geq N(N-1)(l_1+l_2).\label{dtcv4-3}
\end{equation}
Then
\begin{align}
   M_1+M_2 & +2r_c \geq H(Z_1,X_c^{(1,2)})+H(Z_2,X_c^{(2,1)})\nonumber\\
   =&H(Z_1,X_c^{(1,2)},\hat{S}_1^1)+H(Z_2,X_c^{(2,1)},\hat{S}_1^2)\nonumber\\
   =&H(\hat{S}_1^1)+H(Z_1|\hat{S}_1^1)+H(X_c^{(1,2)}|Z_1,\hat{S}_1^1)+H(\hat{S}_1^2)+H(Z_2|\hat{S}_1^2)+H(X_c^{(2,1)}|Z_2,\hat{S}_1^2)\nonumber\\
   \geq&H(\hat{S}_1^1)+H(Z_1|S_1)+H(X_c^{(1,2)}|Z_1,S_1)+H(\hat{S}_1^2)+H(Z_2|S_1)\nonumber\\
   \geq& l_1+2l_2+\frac{N-2}{N-1}\left[H(Z_{2}|S_{1})+H(Z_{1}|S_{1})\right],\label{dtcv5-1}
\end{align}
where (\ref{dtcv5-1}) follows from (\ref{dtcv2-4}).\par
Recall that
\begin{align}
  r_c+M_1
  \geq l_1+l_2+\frac{N-2}{N-1}H(Z_{1}|S_{1})-\frac{1}{N-1} H(Z_{2}|S_{1}).\label{dtcv5-2}
\end{align}
Therefore, by cancelling the term $H(Z_{2}|S_{1})$ in (\ref{dtcv5-1}) and (\ref{dtcv5-2}), we obtain
\begin{align}
  &M_1+M_2+2r_c+(N-2)(r_c+M_1)  \nonumber\\
   \geq& (N-1)l_1+Nl_2+(N-2)H(Z_{1}|S_{1})\nonumber\\
   \geq&(N-1)l_1+Nl_2+\frac{(N-2)(N-1)}{N}H(Z_{1}),\label{dtcv5-3}
\end{align}
where (\ref{dtcv5-3}) follows from (\ref{dtcv1}).\par
Hence, we have
\begin{equation}
  2(N-1)M_{1}+NM_{2}+N^2r_c\geq N(N-1)l_1+N^2l_2.\label{dtcv5-4}
\end{equation}
Similarly, we have
\begin{equation}
  2(N-1)M_{2}+NM_{1}+N^2r_c\geq N(N-1)l_2+N^2l_1.\label{dtcv5-5}
\end{equation}
Finally, from (\ref{dtcv4-2}), (\ref{dtcv4-3}), (\ref{dtcv5-4}), (\ref{dtcv5-5}) and the cut-set bound proved in \cite{yang2016coded}, the converse proof is completed.

\end{appendix}

\bibliography{cache}
\bibliographystyle{IEEEtran}
\end{document}